# Electronic Properties of Electroactive Ferrocenyl-Functionalized MoS$_2$

*Trung Nghia Nguyên Lê,[a] Kirill Kondratenko,[b] Imane Arbouch,[c] Alain Moréac,[d] Jean-Christophe Le Breton,[d] Colin van Dyck,[e] Jérôme Cornil,[c] Dominique Vuillaume[b] and Bruno Fabre[a,\*]*

[a] CNRS, ISCR (Institut des Sciences Chimiques de Rennes)-UMR6226, Univ Rennes, Rennes F-35000, France.

[b] Institute for Electronics Microelectronics and Nanotechnology (IEMN), CNRS, University of Lille, Av. Poincaré, Villeneuve d'Ascq, France.

[c] Laboratory for Chemistry of Novel Materials, University of Mons, B-7000 Mons, Belgium.

[d] CNRS, IPR (Institut de Physique de Rennes)-UMR 6251, Univ Rennes, Rennes F-35000, France.

[e] Theoretical Chemical Physics group, University of Mons, B-7000 Mons, Belgium.

KEYWORDS: Transition metal dichalcogenides (TMDs), MoS$_2$, redox functionalization, ferrocene, charge transport.

ABSTRACT. The attachment of redox-active molecules to transition metal dichalcogenides (TMDs), such as MoS$_2$, constitutes a promising approach for designing electrochemically switchable devices through the control of the material's charge/spin transport properties by the redox state of the grafted molecule and thus the applied electrical potential. In this work, defective plasma treated MoS$_2$ is functionalized by a ferrocene derivative and thoroughly investigated by various characterization techniques, such as Raman, photoluminescence, X-ray photoelectron spectroscopies, atomic force microscopy (AFM) and electrochemistry. Furthermore, in-plane and out-of-plane conductive-AFM measurements (*I–V* and first derivative $\partial I/\partial V$–*V* curves) are measured to investigate the effect of the chemical functionalization of MoS$_2$ on the electron transport properties. While the conduction and valence bands are determined at +0.7 and -1.2 eV with respect of the electrode's Fermi energy for pristine MoS$_2$, additional states in an energy range of ≈ 0.45 eV below the MoS$_2$ conduction band are measured after plasma treatment, attributed to S-vacancies. For ferrocene-functionalized MoS$_2$, the S-vacancy states are no longer observed resulting from the defect healing. However, two bumps at lower voltages in the $\partial I/\partial V$–*V* indicate a contribution to the electron transport through ferrocene HOMO, which is located in the MoS$_2$ band gap at ≈ 0.4/0.6 eV below the Fermi energy. These results are in good agreement with theoretical density functional theory (DFT) calculations and UV photoelectron spectroscopy (UPS) measurements.

## 1. Introduction.

Great advances in the control of graphene electronic properties have stimulated research for new two-dimensional materials with unique electronic and optoelectronic properties.[1] Transition metal dichalcogenides (TMDs), such as molybdenum disulfide MoS$_2$ (and its derivatives MoSe$_2$, MoTe$_2$, WS$_2$, WSe$_2$...), have recently appeared as very promising nanostructures for numerous applications in many areas ranging from electrocatalysis and electronics to optics.[2,3,4,5,6,7,8] Unlike graphene, some TMDs have a semiconductor band gap which make them interesting for applications in electronics, such as high-performance field-effect transistors (FETs) and diodes.[9,10] Moreover, TMDs offer a strong potential for spintronics due to some remarkable properties, such as high spin-orbit coupling (SOC) leading to a unique spin and valley polarization.[11,12,13,14]

The covalent attachment of specific molecular units to these 2D materials in view of tuning their physicochemical properties, enhancing their processability and conferring them novel functionalities is attracting a growing interest.[15,16,17] To date, surface functionalization of TMDs mainly concerned $MoS_2$ with only a few examples devoted to other derivatives.[18,19,20] Among the different functionalization strategies, *e.g.* click chemistry,[21,22] grafting of aryldiazonium salts[23,24] or covalent functionalization using alkyl halides,[25] the reaction of sulfur-containing groups (e.g. organothiols) with metallic 1T-$MoS_2$ (resulting from chemical exfoliation of 2H-$MoS_2$) or semiconducting defect-rich 2H-$MoS_2$ layers has been widely explored for the covalent attachment of molecules via the sulfur vacant sites.[15,17,26,27,28] Through functionalization, it has been demonstrated that the electronic and optical properties of $MoS_2$ could be finely tuned and outstanding effects were reported on the performance of devices integrating such organosulfides functionalized $MoS_2$, such as FETs[29,30] and (bio)sensing platforms.[31,32]

The functionalization of $MoS_2$ with redox-active molecules has surprisingly been much less explored. Yet, such an approach could be highly interesting to design electrochemically switchable electronic and spintronic devices. Indeed, it is anticipated that the charge/spin transport properties in $MoS_2$ could be controlled *in principle* by the redox state of the grafted molecule and thus the applied electrical potential.

As a first step towards these exciting prospects, Zhao *et al.* have nicely reported on the novel ferrocene-functionalized $MoS_2$-based field-effect transistor (FET) architecture.[33] As a matter of fact, the integrated electroactive molecule was responsible for improved FET performance and the tunable doping in the 2D semiconductor. Recent theoretical calculations have validated such an approach by supporting that the transport in MoS$_2$ can be electrically and reversibly switched from an ON to OFF state attributed to the grafted ferrocene.[34] Although ferrocene has been used by other groups to probe and modulate the charge transfer kinetics of MoS$_2$,[35,36,37] to the best of our knowledge, the Samori and co-workers' study[33] constitutes the only experimental example of redox functionalization of MoS$_2$. In their work, 6-(ferrocenyl)hexanethiol was used as the electroactive reagent and was deposited on MoS$_2$ by spin-coating. The FETs measurements were further carried out in a droplet of ionic liquid covering ferrocene-coated MoS$_2$ (MoS$_2$-Fc) material. However, the effects induced by truly bound (chemisorbed) or only physisorbed Fc on the FET transfer characteristics could not be discriminated although control experiments have been carefully achieved by using unsubstituted Fc.

Herein, we report new and complementary findings on the charge transport characteristics of this hybrid material. In our work, the MoS$_2$-Fc material was isolated and its functionalization was thoroughly investigated by various surface characterization techniques including Raman, photoluminescence (PL), X-ray photoelectron (XPS) spectroscopies, and atomic force microscopy (AFM). Furthermore, the effect of the attached electroactive molecule on its charge transport properties was examined by conductive AFM (C-AFM). This technique was used herein to deduce the energy diagram of the MoS$_2$-Fc/electrode junction, which is a prerequisite before employing it in an electrochemically switchable device; the electronic structure of the MoS$_2$-Fc hybrid structure has been further computed at the Density Functional Theory (DFT) level and confronted to the C-AFM data. Detailed electrochemical investigations of these redox-active interfaces, using cyclic voltammetry, provide also a complete view of charge transfer mechanism in these assemblies.

## 2. Experimental

### 2.1. Preparation of MoS$_2$-coated surfaces and further functionalization with ferrocene

*2.1.1. MoS$_2$ exfoliation on SiO$_2$/Si substrates.*

For MoS$_2$ exfoliation, we followed the gold (Au) tape exfoliation method developed by Liu *et al*. (Figure S1 in Supporting Information).[38] Briefly, a 100 nm-thick Au layer was deposited onto a flat SiO$_2$/Si substrate (Graphene Supermarket, $p^{++}$-type Si, 525 $\mu$m thick, 0.001-0.005 Ω·cm coated with a thermal 90 nm-thick SiO$_2$ layer) by thermal evaporation (thin film deposition system) or sputtering. A layer of polyvinylpyrrolidone (PVP) solution (Sigma Aldrich, mw 40000, 10% wt in 1/1 v/v ethanol/acetonitrile) was deposited on the top of the Au surface by spin coating (3000 rpm, acceleration 1000 rpm/s, 2 min) and placed on a hotplate at 150 °C for 5 min. This layer served as a sacrificial layer to minimize the contamination due to the tape residue. The Au/PVP was removed from SiO$_2$/Si with thermal release tape (Graphene Supermarket, release temperature 90 °C), revealing a flat and clean Au surface. The Au/PVP/tape was pressed onto a freshly cleaved bulk of MoS$_2$ (HQ graphene, >99.995%). After lifting off the surface, the tape carried the Au/PVP layer with a monolayer MoS$_2$ crystal attached to Au, and was then transferred onto another SiO$_2$/Si substrate (300 nm thick thermally grown oxide on highly doped $p^{++}$-Si(100) for electrical measurements, from Graphene Supermarket®). The thermal release tape was removed by heating at 200 °C. The PVP layer was then removed by dissolving it in ultra-pure water for 2 h. The Au/MoS$_2$ layer was then rinsed with acetone and cleaned by an air plasma for 3 min (Harrick Plasma cleaner) to remove any remaining polymer residues. The Au layer was dissolved in a KI/I$_2$ (Alfa Aesar) gold etchant solution. Finally, the MoS$_2$ monolayer remaining on SiO$_2$ was rinsed with ultra-pure deionized water (18.2 MΩ cm) and isopropanol, and dried under a N$_2$ flow.

For in plane conductive AFM measurements, Au contacts were deposited by thermal evaporation on the edge of monolayer flakes by lift off after usual photolithography steps.

*2.1.2. MoS$_2$ exfoliation on Au substrates.*

For MoS$_2$ exfoliation on Au, we started from the same Si/SiO$_2$/100 nm-thick Au substrate onto which a drop of optical adhesive (Norland Optical Adhesive 81 (NOA)) was deposited (Figure S2). A glass slide was then pressed on the Au/NOA surface and cured with UV light for 1 h. Subsequently, the Au layer was cleaved with a razor blade and the so exposed Au surface was quickly pressed onto a freshly cleaved MoS$_2$ crystal and heated to 200 °C on a hotplate to improve the MoS$_2$/Au adhesion. After 1 min, the substrate was removed from the hotplate to cool down and the MoS$_2$ bulk crystal was gently detached leaving monolayer flakes on the Au surface with a typical size of a few hundreds of $\mu$m.

*2.1.3. MoS$_2$ exfoliation on SiO$_x$/Si substrates for electrochemistry.*

For cyclic voltammetry measurements, the exfoliation method illustrated in Figure S1 has to be adapted as a low resistance substrate is necessary. We therefore used a $p^{++}$-Si substrate coated with a ca. 1.5 nm-thick native SiO$_x$ layer for the transfer of MoS$_2$ as it enabled further electrochemical measurements to be performed. This substrate was generated by first sonicating $p^{++}$-type (0.001-0.005 $\Omega$ cm resistivity, boron doped, 250 $\mu$m thickness, Siltronix) Si(100) for 10 minutes successively in acetone, ethanol and ultra-pure 18.2 M$\Omega$ cm water and then dipping the so degreased surface into a "piranha" solution (3:1 v/v concentrated H$_2$SO$_4$/30% H$_2$O$_2$) for 30 min at 105 ºC, followed by copious rinsing with ultra-pure water. A 2 nm Cr layer and a 5 nm Au layer were then sputtered on the SiO$_x$/$p^{++}$-Si substrate. Freshly cleaved bulk MoS$_2$ crystal was then deposited at 200 °C on a hotplate with thermal release tape onto the freshly deposited Au surface and removed from the hotplate to cool down. The deposited bulk crystal was then removed from the Au sample surface with a scotch tape leaving large monolayer crystals on Au. To observe solely the electrochemical response of the grafted MoS$_2$ without any potential contributions from molecules adsorbed on the substrate, poly(methyl methacrylate) (PMMA) was spin coated onto the grafted sample, as a protective layer, followed by etching of the metallic layers Au and Cr by $KI/I_2$ gold etchant (30% in water) solution and diluted HCl (10%), respectively. PMMA was then dissolved in acetone leaving Fc-$MoS_2$ crystals on the $p^{++}$-Si(100) surface. For a reference measurement, the same steps were repeated without exfoliated $MoS_2$ crystals on the substrate.

*Caution*: The concentrated aqueous $H_2SO_4/H_2O_2$ (piranha) solution is highly dangerous, particularly in contact with organic materials, and should be handled extremely carefully.

*2.1.4. Plasma treatment of $MoS_2$.*

The creation of defects (sulfur vacancies) on $MoS_2$ was carried out with the $MoS_2$-coated substrate (Au or Si/$SiO_x$) using an air plasma (Harrick Plasma cleaner) for 5 s at a plasma power of 7 W.

*2.1.5. Functionalization of plasma treated $MoS_2$ with ferrocene.*

The plasma treated $MoS_2$-coated substrate was immersed into a 1 mM of 6-(ferrocenyl)hexanethiol (Sigma-Aldrich) solution in acetone (semiconductor grade, Carlo Erba). After degassing with argon for at least 15 min, the substrate was allowed to react with the grafting solution for 24 h at room temperature under argon. In order to remove any excess of the functionalizing agent, it was then subjected to several acetone baths and dried under an argon stream. For each washing cycle, the solvent was collected and subsequently removed under high vacuum. The residues were dissolved again in acetone-$d_6$ and analyzed by $^1$H-NMR spectroscopy (Figure S3).

## 2.2. Instrumentation

*2.2.1. Raman and photoluminescence spectroscopies.* Micro-Raman and micro-photoluminescence spectra were collected with a micro-spectrometer LabRAM HR Evolution800 (Horiba Scientific), respectively with a selected 1800 and 600 g/mm gratings suitable for the spectral resolution required for these studies. Micro-Raman ($\mu$R) measurements have been performed on the 35 to 553 cm$^{-1}$ spectral range (dispersion around 0.51 cm$^{-1}$/pixel), micro-photoluminescence ($\mu$PL) measurements have been performed on the 532 to 869 nm (or 770 nm) spectral range (dispersion around 0.05 nm/pixel) with a laser wavelength of 532 nm. Measurements were acquired by using an Olympus x100 ULWD objective and an optical density filter to avoid laser heating. The XY stage makes $\mu$R and $\mu$PL mappings possible that we used for the calculation of the average spectra of the different domains.

*2.2.2. XPS.* XPS measurements were performed with a NEXSA G2 (ThermoFischer Scientific) spectrometer using the Al K$\alpha$ X-ray source working at 1486.6 eV and using a spot size of 200 $\mu$m². Survey spectra (0-1000 eV) were acquired with an analyzer pass energy of 200 eV (1 eV/step); high-resolution spectra used pass energy of 50 eV (0.1 eV/step). Binding energies were referenced to Au 4f 7/2 peak at 84.0 eV. The core level spectra were peak-fitted using the CasaXPS Software, Ltd. Version 2.3.25PR1.0. Shirley background subtraction was used for the spectral analysis. The peak areas were normalized by the manufacturer-supplied sensitivity factors ($S_{C\ 1s} = 1$, $S_{O\ 1s} = 2.88$, $S_{Mo\ 3d} = 11.0$, $S_{S\ 2s} = 1.29$, $S_{S\ 2p} = 1.88$ and $S_{Fe\ 2p} = 14.35$).

*2.2.3. UPS.* UPS analyses were conducted in an integrated ultrahigh vacuum system, connected to the NEXSA G2 spectrometer (Thermo Fisher Scientific). UPS spectra were obtained with a He lamp providing a resonance line, He I ($h\nu = 21.2$ eV). The spot size was 1mm. The specific used parameters were a dwell time of 50 ms, a step energy of 0.05 eV and a pass energy of 2 eV.

*2.2.4. Electrochemical measurements.* Cyclic voltammetry measurements were performed with an Autolab electrochemical analyzer (PGSTAT 30 potentiostat/galvanostat from Eco Chemie B.V.) equipped with the GPES and FRA softwares in a home-made three-electrode glass cell. The working electrode was $MoS_2/SiO_x/p^{++}$-Si after plasma treatment and ferrocene grafting. It was pressed against an opening in the cell side using a Teflon circular piece and a FETFE (Aldrich) O-ring seal. After an Ohmic contact was made on the previously polished rear side of the sample by applying a drop of In-Ga eutectic (Alfa-Aesar, 99.99%), a steel piece was positioned on the eutectic-coated side of the sample and subsequently the assembly was screwed to the cell using a plastic cap screw. The electrochemically active area of the Si(100) surface (namely 0.38 cm$^2$) was estimated by measuring the charge under the voltammetric peak corresponding to the ferrocene oxidation on an oxide-free, hydrogen-terminated silicon surface and compared to that obtained with a 1 cm$^2$-Pt electrode under the same conditions. A carbon rod and the system $10^{-2}$ M Ag$^+$ | Ag in acetonitrile were used as the counter and reference electrodes, respectively. The potentials versus $10^{-2}$ M Ag$^+$ | Ag can be converted into potentials versus aqueous KCl-saturated calomel electrode (SCE) by adding +0.29 V. Tetra-*n*-butylammonium perchlorate Bu$_4$NClO$_4$ (Fluka, puriss, electrochemical grade) was used at 0.1 mol L$^{-1}$ as supporting electrolyte in acetonitrile. The electrolytic medium was dried over activated, neutral alumina (Merck) for 30 min, while stirring and degassing with argon. About 20 mL of this solution was transferred with a syringe into the electrochemical cell prior to experiments. All electrochemical measurements were carried out inside a home-made Faraday cage, at room temperature (20 ± 2 °C) under constant argon gas flow. The Ohmic drop was determined before each experiment by measuring the impedance of the system at 100 kHz and it was found to be around 500 Ω in (CH$_3$CN + 0.1M Bu$_4$NClO$_4$).

*2.2.5. Atomic Force Microscopy (AFM).* AFM images were acquired on a NT-MDT Ntegra microscope in semi-contact mode with FM tips (APPNANO, SPM Probe Model: ACTA, resonance frequency: 200-400 kHz). The images were treated and analyzed with the open-source Gwyddion software (tilt correction by mean plane corrections).

*2.2.6. Conductive AFM (C-AFM) measurements.* The electron transport properties were measured with a C-AFM (ICON, Bruker) instrument at room temperature (air-conditioned laboratory with controlled $T_{amb}$ = 22 °C and relative humidity of 35-40%) and using an Au coated tip probe with a nominal radius of curvature ≈20 nm. We control the loading force at 12-15 nN. For the in-plane current measurements (field-effect transistor configuration), the tip was grounded, the drain voltage, $V_D$, was applied on the Au lithographed electrode and the gate voltage, $V_G$, on the underlying highly doped Si. The scanned voltage (i.e., $V_D$ for the output *I - V* characteristics, $V_G$ for the transfer *I - V* curves) was applied with the internal voltage source of the ICON C-AFM, while the fixed voltage (i.e., $V_G$ for the output *I - V* characteristics, $V_D$ for the transfer *I - V* curves) was applied by an external Keysight B2987A source/measure unit (SMU). The tip was placed at a distance *L* from the edge of the Au electrode, and 100 *I - V* traces were acquired around this point by moving the tip on a 10 x 10 grid (10 nm pitch) and averaged. For the out-of-plane measurements, we used a "blind" mode to measure the current-voltage (*I - V*) curves and the current histograms: a square grid of $10 \times 10$ was defined with a pitch of 50 to 100 nm. At each point, the *I - V* curve was acquired leading to the measurements of 100 traces per grid. This process was repeated several times at different places (randomly chosen) on the $MoS_2$ sample, and up to several thousands of *I - V* traces were used to build the current-voltage histograms.

The first derivative $\partial I/\partial V$ - *V* curves were numerically calculated from the mean $\bar{I}$ - *V* curves and smoothed with the Savitzky-Golay technique (2nd order polynomial function on 20 data points). For the ferrocene grafted MoS$_2$ junctions, the $\partial I/\partial V$ - $V$ curves were also acquired directly (about 160 curves at different locations on the sample). A small AC voltage at 1 kHz with an amplitude of 30 mV was superimposed to the DC voltage ramp and the output voltage of the C-AFM transimpedance amplifier was fed into the lock-in amplifier to measure $\partial I/\partial V$ (Nanonis Electronics).

## 3. Results and Discussion.

The mechanically exfoliated MoS$_2$ was deposited on either SiO$_2$/$p^{++}$-Si or Au substrate. After deposition, the monolayer MoS$_2$ flakes were visualized by optical microscopy and characterized using Raman and X-ray photoelectron spectroscopies, photoluminescence, atomic force microscopy and electrochemistry. The in-plane and out-of-plane electron transport measurements were performed with MoS$_2$ deposited on SiO$_2$/Si and Au, respectively.

### 3.1. Raman and photoluminescence spectroscopies of different MoS$_2$ samples.

Figure 1 shows the Raman spectra of MoS$_2$ deposited on thermal SiO$_2$-coated Si substrate at different steps. The Raman spectrum of MoS$_2$ typically exhibits two main peaks, the E' and A$_1$' modes, which correspond to in-plane and out-of-plane vibrations of the Mo-S bond, respectively. For pristine MoS$_2$, the E' and A$_1$' peaks are located at 385.83 and 404.21 cm$^{-1}$, respectively. The wavenumber difference of 18.38 cm$^{-1}$ between the two is consistent with MoS$_2$ monolayer.[39,40] After plasma treatment for 5 s, the E' peak shifted towards lower wavenumbers (redshift) whereas the A$_1$' peak shifted towards higher wavenumbers (blueshift). Alongside with these shifts, we observe the appearance of shoulders on the low (high) frequency side of the E' (A$_1$') peaks as already observed for defective MoS$_2$.[41,42] The Raman spectrum after plasma treatment is also characterized by the appearance of disorder-induced peaks in the low frequency region (120-260 cm$^{-1}$). Among these, a prominent peak at ~227 cm$^{-1}$ has been assigned to the longitudinal acoustic phonons with momentum q≠0 at the M point labelled LA(M) that can be used to quantify the defect density (*vide infra*).[41,43]

After grafting of ferrocene on defective MoS$_2$, we observe a strong blueshift of the E' peak with a frequency similar to that of the pristine sample, a redshift of the A$_1$' peak, and the reduction of the defect-activated mode intensity. This intensity reduction is attributed to the "healing" of S-vacancy sites.[15,33,44,45,46] To verify the influence of the presence of S-vacancy sites on the grafting efficiency, a control experiment was carried out for which the pristine MoS$_2$-coated surface was immersed in the 6-(ferrocenyl)hexanethiol solution for 24 h. The similarity of Raman spectra before and after immersion in the grafting solution highlights the importance of defects for efficient grafting of MoS$_2$. It is also indicative of the very low density of defect sites in as-prepared pristine MoS$_2$ and hence a low number of grafting sites (Figure S4).

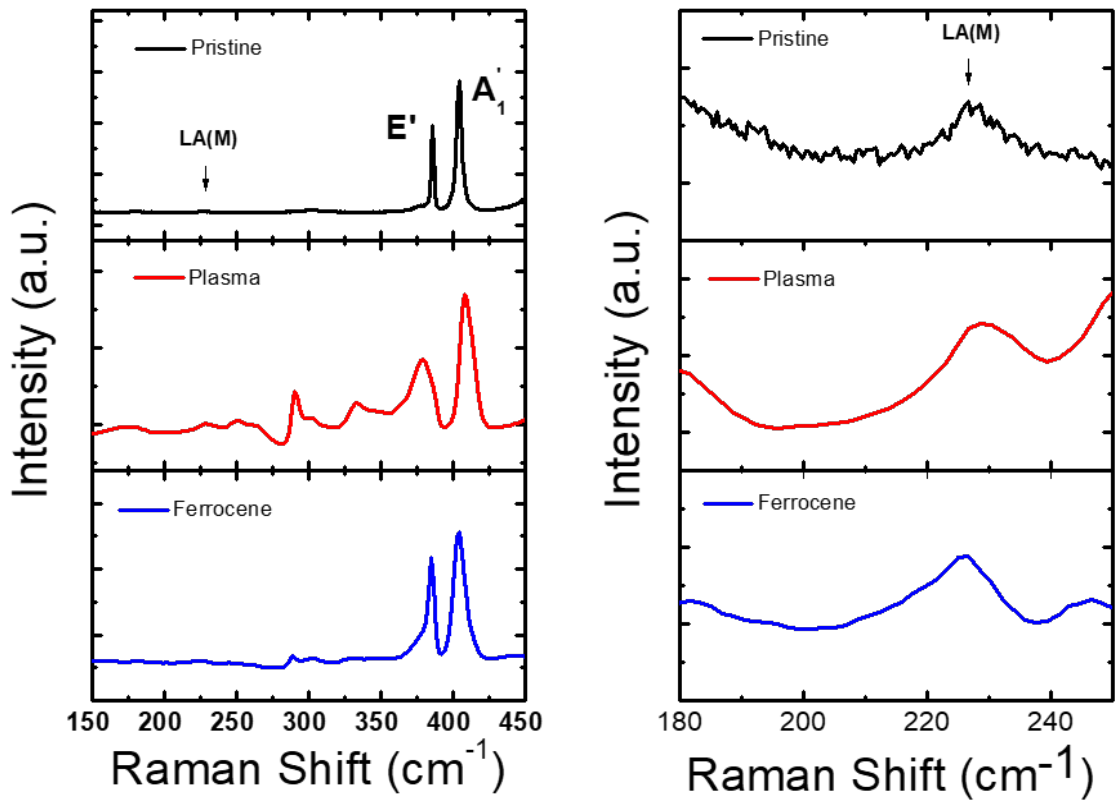

**Figure 1.** (Left) Raman spectra of pristine, plasma-treated and ferrocene-grafted MoS$_2$ deposited on SiO$_2$/Si. (Right) Zoom on the LA(M) peak.

The intensity of defect-activated Raman peaks is utilized to quantify the defect density (Figure S5). Indeed, it was demonstrated that the ratio between the intensities of the LA(M) and the $A_1'$ peaks is inversely proportional to the square of the mean inter-defect distance in the MoS$_2$ sheet through the relationship $\frac{I(\text{LA})}{I(A'_1)} = \frac{\gamma}{L_D^2}$ where $L_D$ is the mean inter-defect distance and $\gamma$ is a correlation constant.[41,43] In the absence of a proper measurement of the number of defects created by the plasma treatment, the value of the correlation constant was based on the literature. Using the value of $\gamma = 0.59$ nm² determined by Mignuzzi et al.,[41] a mean inter-defect distance of 3.1 nm was calculated. This value can be converted into a defect density $n_D$ of $3.3 \times 10^{12}$ cm$^{-2}$ using the following expression $n_D(\text{cm}^{-2}) = 10^{14}/(\pi L_D^2)$[47] (Table 1). After grafting, the $\frac{I(\text{LA})}{I(A'_1)}$ intensity ratio is strongly reduced which corresponds to an increased mean inter-defect distance of 11.9 nm. This value is higher than that determined after plasma treatment and is consistent with a healing effect of the grafting solution. This interdefect distance corresponds to a defect density of 2.2 x 10$^{11}$ cm$^{-2}$. Assigning this reduction solely to the thiol grafting on defect sites would result in 3.1 x 10$^{12}$ grafted molecules cm$^{-2}$ (93% of the defect sites). Note that the inter-defect distance is strongly dependent on the correlation constant taken from the literature (an order of magnitude higher correlation constant was determined by Aryeetey et al.[43]) and we can not rule out a possible healing effect from the oxygen-containing solvent molecules, as previously reported by other groups.[46,48,49] Nevertheless, the use of acetone as the thiol grafting solution instead of commonly employed ethanol is thought to restrict this parasitic reaction to a certain extent.

**Table 1.** Determination of the defect density of pristine, plasma-treated and ferrocene-grafted MoS$_2$ deposited on SiO$_2$/Si from Raman spectroscopy data.

| MoS$_2$ sample | $I$(LA)/$I$(A'$_1$) | $L_d$ (nm) | Defect density (cm$^{-2}$) |
|---|---|---|---|
| **Pristine** | 0.040 | 18.9 | $8.8 \times 10^{10}$ |
| **Plasma** | 0.247 | 3.1 | $3.3 \times 10^{12}$ |
| **Ferrocene** | 0.092 | 11.9 | $2.2 \times 10^{11}$ |

In addition to Raman spectroscopy, photoluminescence (PL) is a powerful tool to characterize the electronic properties of TMDs. The PL of MoS$_2$ is extremely sensitive to strain[50,51,52] or to the modification of the charge doping in the layer upon molecular adsorption[33,46,53,54] or back gate voltage.[55,56] The PL spectrum of pristine MoS$_2$ can be decomposed into three main components which can be ascribed to neutral exciton A(X$^0$), negatively charged exciton or trion (A(X$^-$)) and B exciton.[55] Pristine MoS$_2$ is intrinsically *n*-doped with a PL spectrum dominated by the X$^-$ component. The evolution of the spectral weight of the trion is an indication of the charge density change upon grafting.[46,53,54] Figure 2 displays the PL spectra of monolayer MoS$_2$ deposited on SiO$_2$ for the pristine sample, after plasma treatment and further ferrocene grafting. Pristine MoS$_2$ shows a PL spectrum dominated by the X$^-$ component, consistent with *n*-type doping. After plasma treatment, we observe a strong increase of the PL intensity as well as a shift of the overall spectrum towards higher photon energy. The intensity ratio between the trion and the neutral exciton associated peaks is strongly reduced. After ferrocene grafting, we observe a strong redshift of the PL spectrum, a decrease of the PL with an increased A(X$^-$)/[A(X$^-$) + A(X$^0$)] intensity ratio indicating an excess of negative charges in the sheet compared to the plasma-treated sample (Table 2).

For the PL of pristine MoS$_2$, the large spectral weight (0.77) of the trion peak is an indication of an excess of electrons in the conduction band and therefore of *n*-type doping of the pristine material.[55] After plasma treatment, the PL intensity increases and the blueshift of the overall spectrum has been interpreted as a *p*-type doping induced by the adsorption of molecules (such as $O_2$ and $H_2O$) from the ambient environment that preferentially adsorb on S-vacancy sites.[57,58] After ferrocene grafting, the shift of the PL peaks towards lower energy and the decrease of the PL intensity are consistent with both a healing effect and a *n*-type doping induced by the attached ferrocene molecule. At this stage, the trion spectral weight increases to 0.87, which is even higher than that determined for the pristine sample demonstrating a pronounced *n*-type doping after the grafting step.

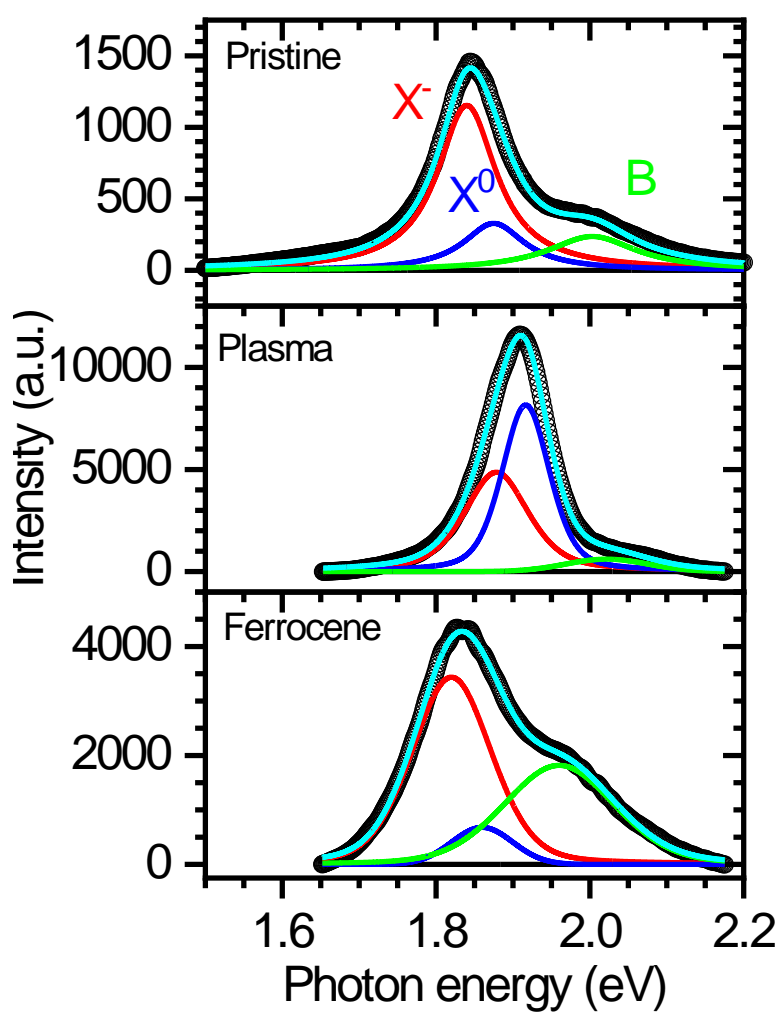

**Figure 2.** Photoluminescence spectra of pristine, plasma-treated and ferrocene-grafted $MoS_2$ deposited on $SiO_2/Si$.

**Table 2.** Integrated intensities of the X$^-$ and X$^0$ PL peaks and spectral weight of the X$^-$ peak for the different MoS$_2$ samples deposited on SiO$_2$/Si.

| MoS$_2$ sample | $I$ (A(X$^-$)) | $I$ (A(X$^0$)) | $I$ (A(X$^-$))/[$I$ (A(X$^-$)) + $I$ (A(X$^0$))] |
|:---:|:---:|:---:|:---:|
| **Pristine** | 158 | 46 | 0.77 |
| **Plasma** | 588 | 722 | 0.44 |
| **Ferrocene** | 470 | 67 | 0.87 |

## 3.2. XPS and AFM.

The grafting of the ferrocene-terminated chains onto plasma treated MoS$_2$ was confirmed by X-ray photoelectron spectroscopy (XPS). The high-resolution Mo 3d spectra of the three MoS$_2$-coated surfaces show the main Mo 3d$_{5/2}$ and Mo 3d$_{3/2}$ components attributed to the lattice Mo(IV)-S at 229.5 ± 0.1 eV and 232.6 ± 0.1 eV, respectively (Figure 3 a-c).[59,60,61,62] For all samples, MoS$_2$ was partially oxidized to MoO$_3$, as supported by the presence of the components observed at higher binding energies, i.e. 232.6 ± 0.1 eV (Mo 3d$_{5/2}$) and 235.7 ± 0.1 eV (Mo 3d$_{3/2}$)[61,62]. From the area of the Mo 3d$_{3/2}$ peak, the atomic ratio Mo(VI)/Mo(IV) was estimated to be less than 13%. Moreover, the atomic concentrations of Mo(VI) of 4.0, 3.8 and 3.4 % determined for the pristine, plasma-treated and ferrocene-grafted samples, respectively, indicate that the oxidation level of the material was not significantly changed after the plasma treatment and the molecular grafting steps. The S 2s signal appears at 226.7 ± 0.1 eV for the three samples, in good agreement with sulfur in MoS$_2$.[61,62] The S 2p spectra display the characteristic spin-split doublets, 2p$_{3/2}$ and 2p$_{1/2}$, with an energy splitting of 1.2 eV and an intensity ratio of 1:2, as theoretically expected (Figure 3 d-f).[59,61] For all samples, the peaks at 162.4 eV and 163.6 eV are assigned to S 2p$_{3/2}$ and S 2p$_{1/2}$, respectively. The experimental ratio between the areas under the S 2s and Mo 3d peaks decreases by about 3% after the plasma treatment and increases by 1.4% to a value closer to that measured for the pristine MoS$_2$ after ferrocene grafting. This trend is consistent with a first step leading to the loss of sulfur atoms followed by their replacement by the thiol-terminated chains. As expected, the two-spin orbit split components p$_{3/2}$ and p$_{1/2}$ of Fe 2p are detected only for the ferrocene functionalized MoS$_2$ sample. Both Fe(II)- and Fe(III) components are observed, with a Fe(III)/Fe(II) ratio of 4.9, at 707.8 and 720.5 eV as well as 710.2 and 723.6 eV, respectively (Figure 3 g-i). The ferrocenium Fe(III) species[63] which are detected at higher binding energies, may result from the internal charge transfer between the ferrocenyl groups and the highly reactive, plasma-treated MoS$_2$, as previously observed for ferrocene intercalated MoS$_2$.[36] As demonstrated in the next section, these ferrocenium units can be reversibly converted to their corresponding reduced ferrocene species by electrochemistry. Two other components located at 714.7 (Fe 2p$_{3/2}$) and 728.4 eV (Fe 2p$_{1/2)}$ have been used to fit the Fe 2p spectrum and correspond to shake-up satellite peaks.[64,65]

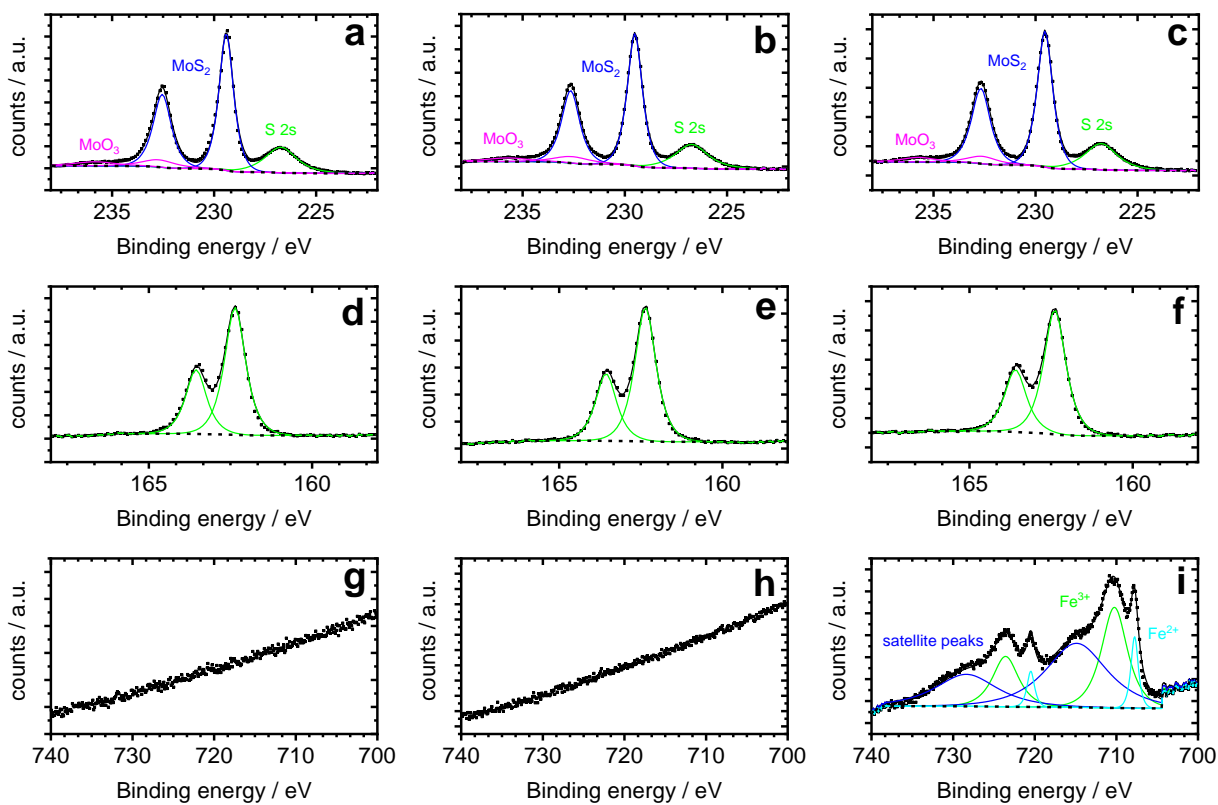

**Figure 3.** High-resolution XPS spectra for the Mo 3d (a-c), S 2p (d-f) and Fe 2p (g-i) peaks of pristine (a, d, g), plasma-treated (b, e, h) and ferrocene-grafted (c, f, i) MoS$_2$ deposited on SiO$_2$/Si. Experimental data and fitting envelopes are represented by black dotted and solid lines, respectively. The colored lines are fitted curves using Gaussian−Lorentzian mixed peaks corresponding to different components.

The surface morphology of MoS$_2$ deposited on SiO$_2$/Si at each step was assessed by AFM. The AFM images of the pristine and plasma treated MoS$_2$-modified surfaces show a homogeneous and smooth surface without apparent features (Figures 4 and S6). The measured root-mean-square (rms) roughness slightly increased from 1.9 to 2.6 Å after plasma treatment, as a result of increased strain caused by the creation of the sulfur vacancy sites. For the thickness measurements, we have focused on the step edges of the MoS$_2$ flakes, measuring the height profiles with respect to the underlying SiO$_2$ substrate (Figure 4). The thickness of the MoS$_2$ layer is estimated to be 1.3 ± 0.2 nm and 0.9 ± 0.1 nm for the pristine and plasma-treated $MoS_2$ samples, respectively, in close agreement with the expected value for one $MoS_2$ monolayer.[60,66,67] After the ferrocene grafting, the surface is also globally compact but numerous homogeneously distributed cracks are visible (Figure 4c). The origin of these cracks is unclear at the moment but we believe that they could originate from strong strain and steric constraints caused by the anchoring of ferrocene-terminated chains. The rms roughness in the cracks-free areas is comparable to that of the plasma treated $MoS_2$. Upon the ferrocene grafting, a ca. 1.4 nm height increase is observed which is perfectly consistent with the theoretical length of the ferrocene-terminated alkyl chain considering the length of the hexyl bridge (7.8 Å)[68] and the diameter of the ferrocene molecule (6.6 Å).[69]

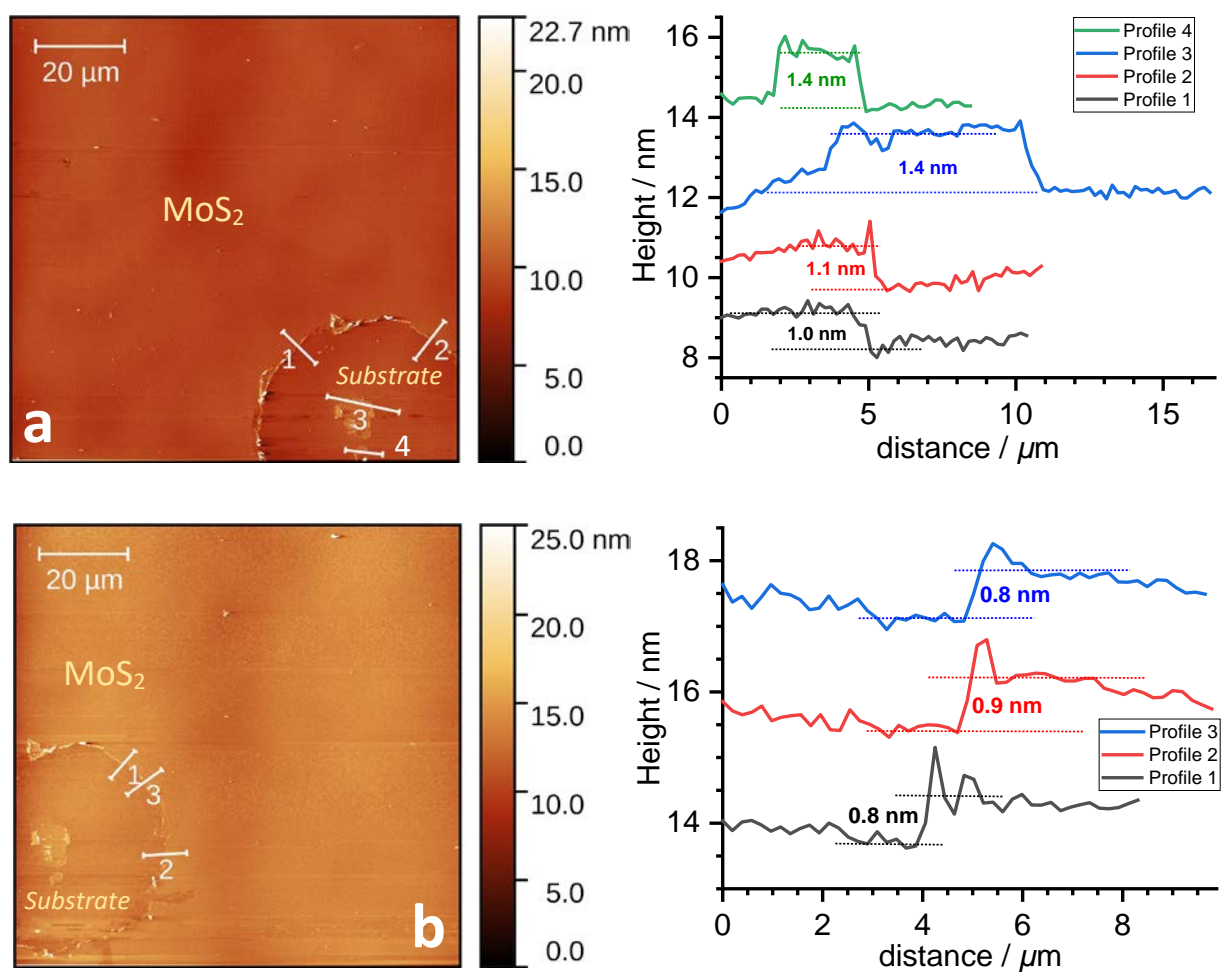

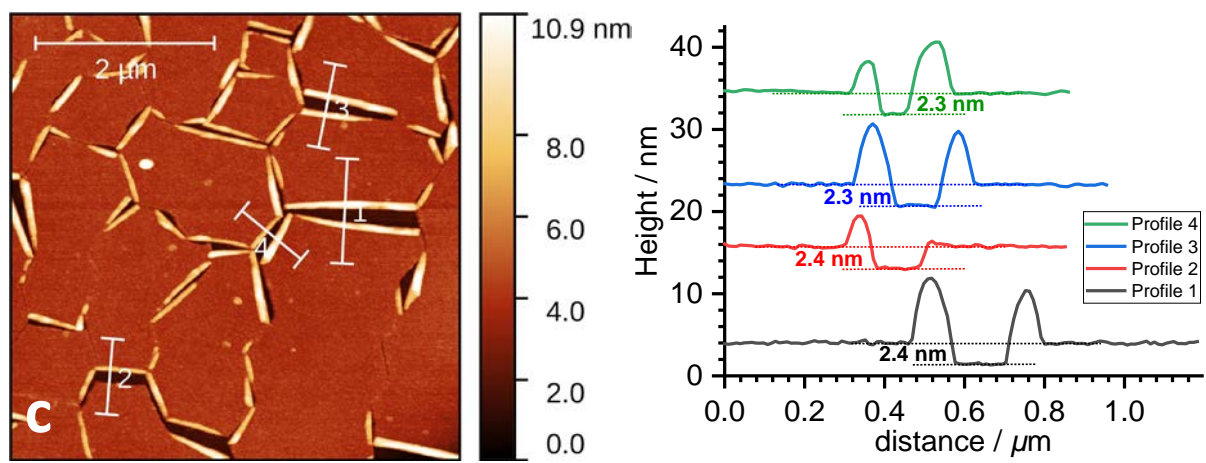

**Figure 4.** AFM images of pristine (a), plasma-treated (b) and ferrocene-grafted (c) MoS$_2$ deposited on SiO$_2$/Si. (Right) Corresponding cross-section profiles taken along the lines in the left images.

### 3.3. Electrochemistry.

Typical cyclic voltammograms of the ferrocene-grafted MoS$_2$ monolayer in CH$_3$CN + 0.1 M Bu$_4$NClO$_4$ are shown in Figure 5 as a function of the potential scan rate $v$. The reversible one-electron wave of the ferrocene/ferrocenium couple is clearly visible at $E^{\circ\prime} = 0.08 \pm 0.01$ V vs 10$^{-2}$ M Ag$^+$/Ag (i.e. 0.37 V vs SCE), in good agreement with the $E^{\circ\prime}$ of dissolved ferrocene measured at a platinum electrode (0.09 V vs 10$^{-2}$ M Ag$^+$/Ag, Figure S7). In addition, good stability of the electrochemical response was observed over the course of 10 successive cycles (Figure S8) and, as expected for surface-confined reversible redox species, the ratio of anodic to cathodic peak current densities ($j_{pa}/j_{pc}$) is close to unity and the peak currents are proportional to the potential scan rate (Figure 5b).[70] It is however noteworthy that the peak-to-peak separation $\Delta E_p$ close to 50 mV is higher than the theoretically expected value of 0 mV, which can be reasonably explained by electron transfer kinetic limitations due to the insulating SiO$_x$ layer between the heavily doped Si and MoS$_2$, and possibly to the semiconducting character of the MoS$_2$ layer (Figure S7b,c). Moreover, to ensure that the observed reversible redox system is solely attributed to ferrocene grafted on MoS$_2$, the washing procedure was monitored by $^1$H-NMR spectroscopy and, after three washing cycles, no traces of the grafting molecule could be detected (Figure S3) indicating the removal of the majority of the physisorbed species. Furthermore, a control experiment was conducted consisting of immersing the surface without $MoS_2$ into the thiol grafting solution. As expected, no redox response is observed which confirms the selective attachment of ferrocenyl units to defective $MoS_2$ (Figure S9).

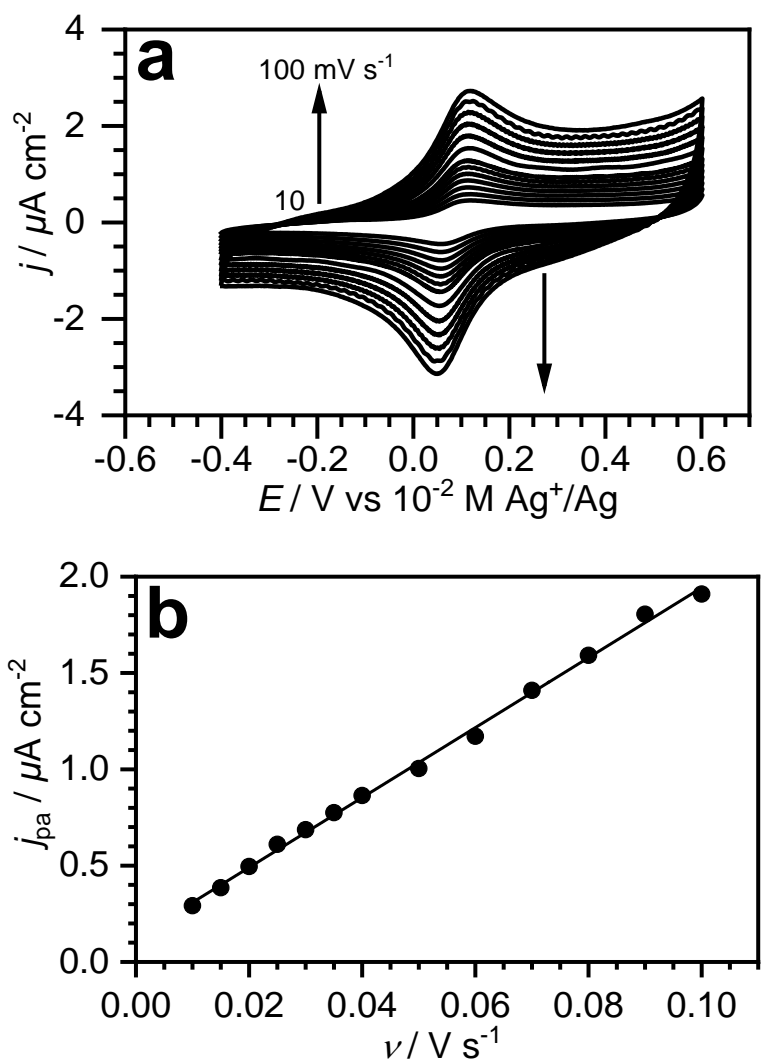

**Figure 5.** a) Cyclic voltammograms at different scan rates between 10 and 100 mV s$^{-1}$ of ferrocene-functionalized $MoS_2$ deposited on $SiO_x/Si$ in $CH_3CN$ + 0.1 M $Bu_4NClO_4$. b) Corresponding plot of the anodic peak current density $j_{pa}$ versus the potential scan rate $v$.

## 3.4. Electronic transport measurements.

*3.4.1. In-plane transport.*

The in-plane electronic transport (ET) was measured in a field-effect transistor (FET) configuration (the FET fabrication is detailed in the Experimental Section). The MoS$_2$ flake was connected between a lithographed Au electrode (the drain on which the voltage was applied) and the Au C-AFM tip (the source, grounded) placed on the flake at a distance *L* (channel length) from the drain electrode. The underlying substrate was 300 nm thick SiO$_2$/$p^{++}$-Si(100), the doped Si being used as the gate electrode. Figure 6a shows two typical transfer characteristics (drain current vs. gate voltage, $I_D$ - $V_G$) in the linear regime (low drain voltage $|V_D| < |V_G|$) for the pristine MoS$_2$ and ferrocene-grafted MoS$_2$ (the current is the mean current calculated from 100 individual *I* - *V* traces). A decrease of the effective electron mobility, $\mu_{eff}$, by a factor of about 10 was observed from 2.8 - 22.8 cm$^2$ V$^{-1}$ s$^{-1}$ (pristine) to 0.35 - 2.6 cm$^2$ V$^{-1}$ s$^{-1}$ (ferrocene-grafted) on several samples (Fig. 6b, Table S1, $\mu_{eff}$ is calculated from the transconductance $\partial I_D/\partial V_G$, considering an effective width *W* for the MoS$_2$ transistor channel, see Section 7 in Supporting Information). Albeit the ill-defined source-drain geometry (C-AFM contact) and the lack of performance optimization through device/technology engineering (e.g. high-k dielectric, double gate, etc…),[71,72] this effective room-temperature mobility (likely overestimated, see discussion in section 7 of the SI) is consistent with the reported values (0.1-10 cm$^2$ V$^{-1}$ s$^{-1}$)[72,73] for transistor with exfoliated single MoS$_2$ layer on SiO$_2$.

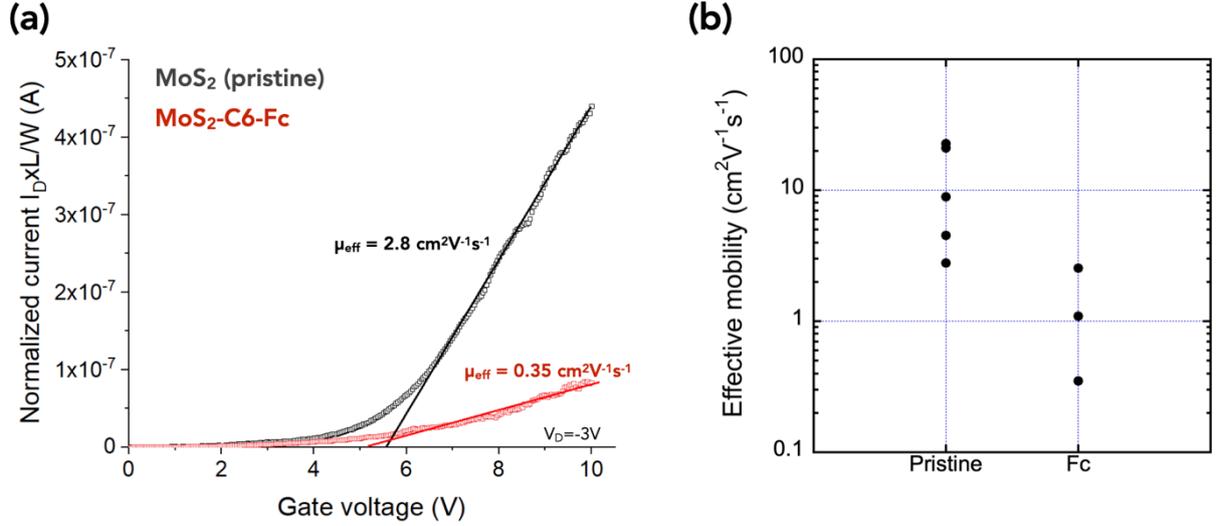

**Figure 6**. (a) Normalized drain current, $I_D$ $L/W$ versus gate voltage $V_G$ for pristine and ferrocene-grafted $MoS_2$ samples. We plot a normalized drain current for the purpose of comparison since the exact values of $L$ and $W$ depend on (*i*) the C-AFM tip position on the $MoS_2$ flake (not strictly constant from sample-to-sample), (*ii*) the size of the $MoS_2$ flakes and (*iii*) their contact with the Au lithographed electrode (see details in the Sections 8 and 9 of the Supporting Information). The mobility is calculated from the slope of the linear part (solid lines), the intercept with the x-axis gives the threshold voltage (≈ 5-5.5 V). (b) Summary of the calculated effective mobility for the different devices (dataset in Table S1).

Figure 7 shows the in-plane current versus $V_D$ at a fixed $V_G$. The *I - V* curves are strongly asymmetric because of the strong geometrical asymmetry between the lithographed Au electrode (hundreds of $\mu$m) and the tiny C-AFM tip (typically tens of nm). We used a double Schottky barrier model to analyze these *I - V* curves and to estimate the Schottky barrier height (SBH) at both interfaces.[74,75,76] This model considers two back-to-back Schottky diodes (Figure 7a), one for the Au electrode/$MoS_2$ interface (SB1) and one for the $MoS_2$/C-AFM tip interface (SB2). The measured current is always limited by the saturation current of the reverse-biased diode, the other diode being in the forward regime (inset in Fig. 7b). Thus, the Au electrode/$MoS_2$ diode SB1 (with a SBH $\phi_{B1}$) is measured at $V < 0$ while the MoS$_2$/Au C-AFM tip diode SB2 (with a SBH $\phi_{B2}$) is measured at $V > 0$.

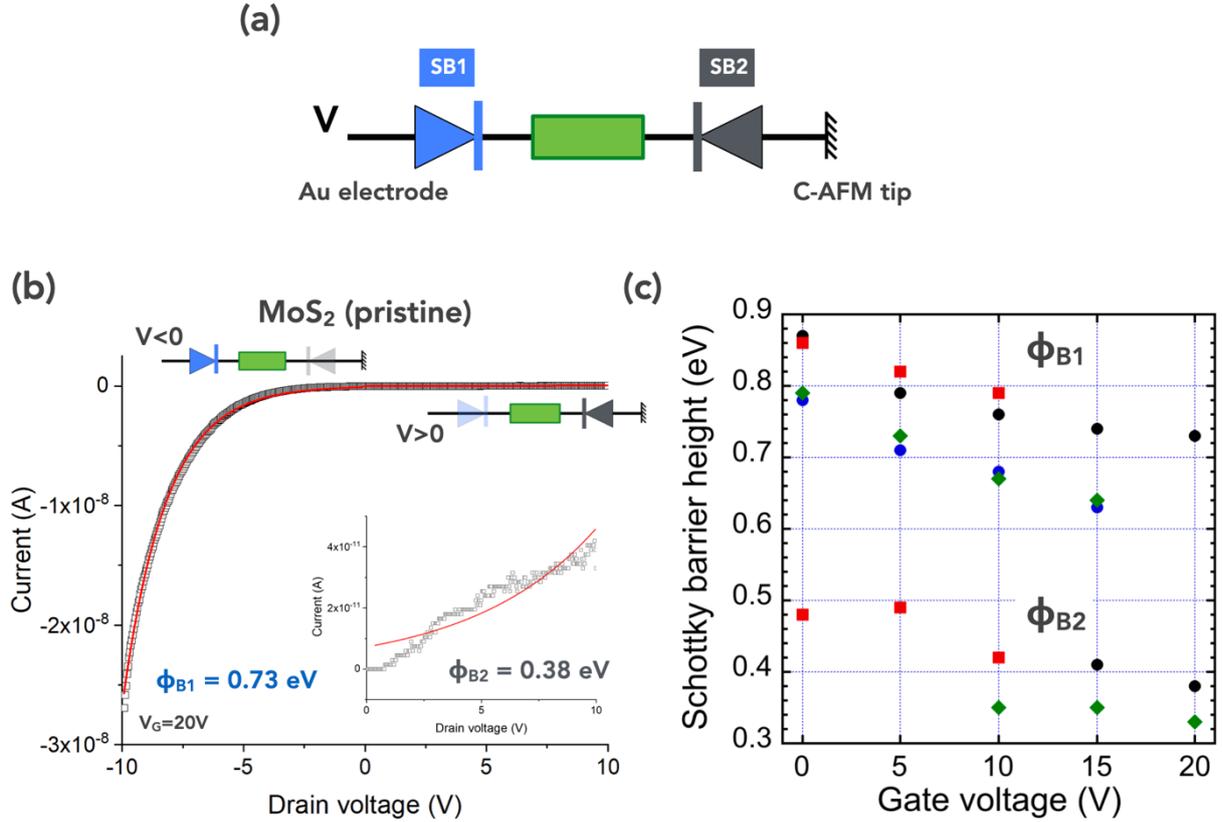

**Figure 7**. (a) Equivalent electrical circuit for the in-plane ET of the Au electrode/MoS$_2$/C-AFM tip device. (b) Typical in-plane current vs. drain voltage for the pristine MoS$_2$ sample (black curve) and fits with eq. 1 (red line). The inset shows the $V > 0$ part. The extracted SBHs $\phi_{B1}$ and $\phi_{B2}$ are marked in the panel. In the equivalent electrical scheme, the diode in the forward regime at $V < 0$ and $V > 0$, respectively, is shaded. (c) SBHs measured for pristine (sample #1, black disks; sample #2, blue disks); plasma-treated (green diamonds) and ferrocene-grafted (red squares) MoS$_2$ samples. The values for some samples were not obtained at all V$_G$ since in several cases, the current was either below the sensitivity limit of the C-AFM preamplifier or very noisy.

The Schottky diode current in saturation is voltage dependent due to interface defects or contaminants and image-force lowering that induce a SBH dependent on the applied voltage. It is expressed by:[77,78]

$$I_{S1,2} = \pm S_{1,2} A^* T^2 exp\left(-\frac{\phi_{B1,2} - q|V|\left(1 - \frac{1}{n_{1,2}}\right)}{kT}\right) \quad (1)$$

with $q$ the electron charge (1.602 ×10$^{-19}$ C), $V$ the applied bias, $k$ the Boltzmann constant (1.38 × 10$^{-23}$ J K$^{-1}$), $T$ the temperature (295 K), $n_{1,2}$ the ideality factors, $S_{1,2}$ the contact areas of the junctions, $A^*$ the Richardson constant ($A^* \approx 5 \times 10^5$ A m$^{-2}$ K² for a MoS$_2$ monolayer considering an electron effective mass of 0.35 - 0.41 $m_0$ for a MoS$_2$ monolayer)[79,80,81], $\phi_{B1}$ and $\phi_{B2}$ the SBHs at zero bias. $S_1$ corresponds to the contact area of the MoS$_2$ flake with the lithographed electrode (estimated between 3 x 10$^3$ and 2 x 10$^4$ $\mu$m² from optical images, see section 8 and Figure S10 in the Supporting Information) and $S_2$ is the contact surface of the C-AFM tip on the flake, estimated at around 10 nm² (Section 9, Supporting Information). The values of $\phi_{B1}$ and $\phi_{B2}$ are given by the fit of eq. (1) on the $V < 0$ and $V > 0$ parts of the $I$ - $V$ curves, respectively, and an example is shown in Figure 7b for the pristine MoS$_2$ sample (and in Figure S11 for the plasma-treated and ferrocene-grafted MoS$_2$ samples). The SBHs measured versus the applied gate voltage are given for the different samples in Figure 7c. Finally, in all the cases, the ideality factors are $\approx 1.10 \pm 0.05$ without any significant trend dependent on the nature of the samples.

Globally, our results can be rationalized as follows. The decrease of the mobility after the ferrocene grafting is in agreement with the report from Zhao *et al.*[33] The large dispersion of the mobility may be due to the flake-to-flake variability and/or variability of the contact resistance at both the lithographed Au electrode and the C-AFM tip. The SBHs at the Au electrode/MoS$_2$ interface ($\phi_{B1}$) and the gate voltage induced SBH lowering (Figure 7c, mainly for $\phi_{B1}$) are in agreement with the literature for transistor measurements.[82,83] The SBHs at the MoS$_2$/Au C-AFM tip are systematically lower ($\phi_{B2} < \phi_{B1}$) by about 0.3 eV. This feature is reasonably explained by considering that the work function of a nanoscale metal-coated C-AFM tip (*ca.* 4.8 eV) is smaller than the one of the "bulk" material (*ca.* 5.1-5.4 eV),[84,85,86] giving a lower energy barrier. Finally, regarding the dispersion of the SBHs ($\approx$0.1 eV between the data for the two pristine $MoS_2$ samples, Fig. 7c), we do not see any significant differences in the SBHs after the ferrocene grafting in our transistor configuration with a C-AFM tip as a movable electrode on the top of $MoS_2$ and a lithographed bottom drain electrode.

*3.4.2. Out-of-plane transport.*

The out-of-plane electron transport characteristics measured by C-AFM through the $MoS_2$ samples deposited on Au substrate are displayed in Figure 8. The 2D histograms of the absolute value of the current (in a log scale) versus the applied voltage (heat map) were constructed from 200 individual *I - V* traces (no data selection). For the pristine $MoS_2$, we clearly observe two groups of *I - V* traces (Figure 8a), as also evidenced by the double log-normal distribution of the current (at -1 V) in Figure 8d. The group with a very low current ($\approx$ 34% of the dataset, mainly at the sensitivity limit of the C-AFM amplifier, *i.e.* < $10^{-10}$ A) is likely due to a poor electrical contact at the C-AFM tip/sample interface resulting from the variability of the loading force, changes in the tip geometry, the contact area and the tip/sample interactions. The mean current - voltage curve ($\bar{I}$ - *V*, black line in Figure 8a) was calculated for the second group (log $|\bar{I}|$ = -7.83 ± 0.46 at -1 V, Figure 8d). After the plasma treatment (Figure 8b), the current significantly increases, especially at low voltages (between -1 and +1 V). The distribution of the current at -1 V is fitted by a log-normal function with a larger standard deviation (log$|\bar{I}|$ = -7.3 ± 1.0, Figure 8e), albeit a less dispersed subgroup ($\approx$ 35% of the dataset) can be distinguished on the 2D histogram (Figure 8b) and the current distribution (marked by a star in Figure 8e). This large distribution of the *I - V* traces can be induced by some inhomogeneities of the plasma treatment, *i.e.* inhomogeneous density of S-vacancy defects. After the ferrocene grafting (Figure 8c), the current strongly decreases (73% of the dataset with log $|\bar{I}|$ = -9.36 ± 0.77 at -1 V, Figure 8f) as expected since the thickness of the junction has increased by ≈ 1.4 nm (as estimated by AFM, *vide supra*). However, we also observe another group (27% of the dataset) with almost the same level of current (log$|\bar{I}|$ = -7.00 ± 0.36 at -1 V) as that measured for the plasma-treated $MoS_2$ sample. This feature is rationalized by assuming that some local zones of the defective $MoS_2$ surface are not functionalized with the ferrocenyl molecules, probably due to some inhomogeneity of the S-vacancies density after the plasma treatment (*vide supra*).

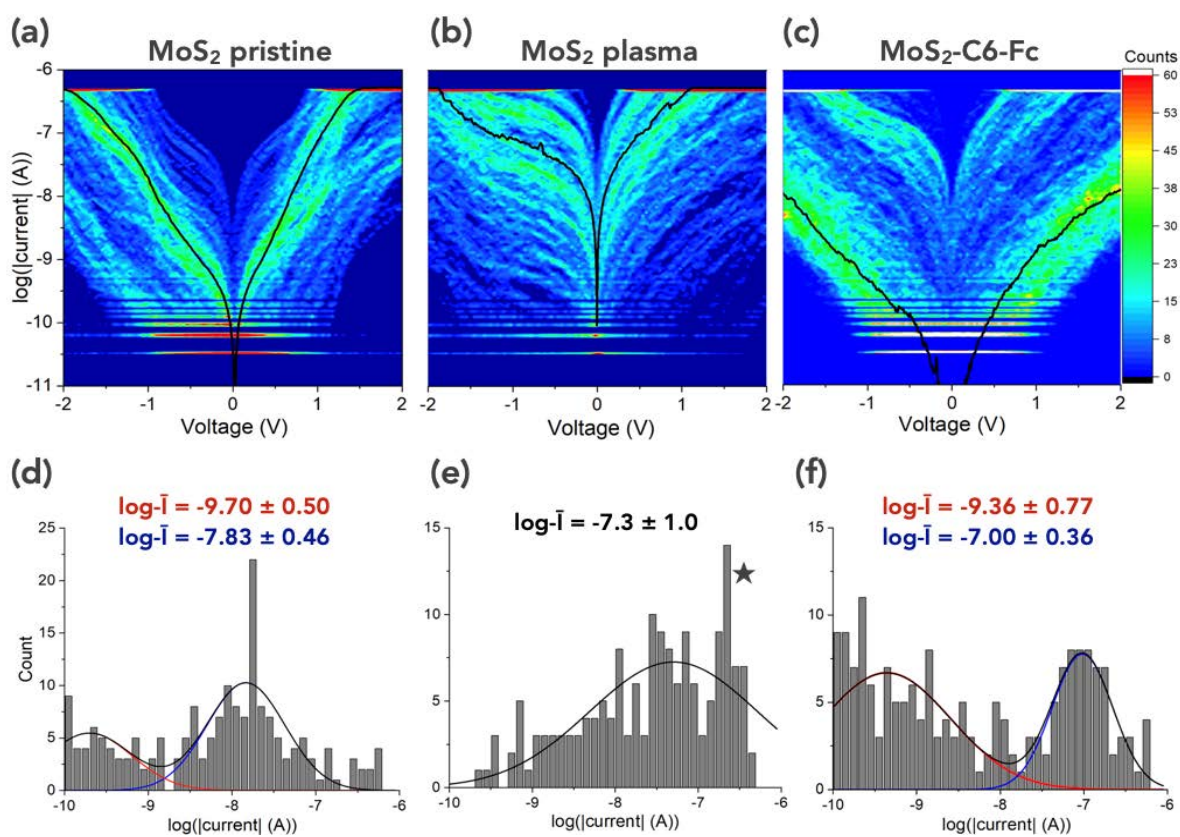

**Figure 8**. 2D histograms (log($|I|$) - *V*, 200 *I* - *V* traces) for (a) the Au/pristine $MoS_2$/Au C-AFM, (b) the Au/plasma-treated $MoS_2$/Au C-AFM and (c) the Au/ferrocene-grafted $MoS_2$/Au C-AFM tip junctions. The three panels share the same y-axis scale and the count scale. The horizontal (red or white) lines at low (< $10^{-10}$ A) and high currents (at 5 x $10^{-7}$ A) correspond to *I* - *V*s in the dataset that reached the sensitivity limit of the C-AFM equipment (almost flat and noisy *I* - *V*

traces with random staircase behavior) or that reached the saturating current compliance of the C-AFM preamplifier during the voltage scan. The panels (d-f) show the corresponding statistical distribution of the current at -1 V and the log-normal fits (solid lines) enabling the log|mean current| ($\log|\bar{I}|$) and standard deviation to be determined.

The first derivative $\partial I/\partial V$ of the $I – V$ curves has been plotted for the three samples to deduce the energy diagram of the junctions (Figure 9). For the pristine MoS$_2$ sample (Figure 9a), we can determine two onset voltages at the intercept between the linear extrapolations of the two branches of the $\partial I/\partial V – V$ spectrum and the background (dashed red lines): $V_{O+} \approx 0.7$ V and $V_{O-} \approx -1.2$ V. Due to the Fermi level pinning at the metal/MoS$_2$ interface,[82,87,88] the energy positions of the MoS$_2$ conduction band (CB) and valence band (VB) with respect to the Au substrate Fermi energy are not shifted by the applied voltage (energy diagram in Figure 9a). The voltage drop is mainly located at the MoS$_2$/C-AFM tip interface. Thus, we can conclude that $V_{O+} \approx 0.7$ V corresponds to the situation when the bottom of the CB enters in the energy window defined by the difference of the Fermi levels of the two electrodes and that electrons can be transferred though the MoS$_2$ CB (while only tunneling is possible below $V_{O+}$). We estimate that the energy barrier with the CB is $\approx 0.7$ eV. For the same reasons, at the negative voltage, from $V_{O-}$, the energy barrier for the VB is $\approx 1.2$ eV, leading to a MoS$_2$ band gap of $\approx 1.9$ eV. All these values are in good agreement with previously reported results for a monolayer of MoS$_2$ on Au electrodes from electrical measurements, DFT calculations,[89,90,91,92] and recently published angle-resolved UV photoelectron spectroscopy (UPS)/inverse photoelectron spectroscopy (IPES) measurements.[93] Our UPS measurements point to an energy position of the MoS$_2$ VB at 1.34 eV below the Fermi energy for the pristine MoS$_2$ (Figure S13) and 1.13 eV for the plasma-treated MoS$_2$, in reasonable agreement with the above determined values from the electrical measurements. After the plasma treatment, the main feature is a reduction of $V_{O+}$ (Figure 9b) due to the creation of S-vacancy states in an energy range of ≈ 0.45 eV below the MoS$_2$ CB, in agreement with a similar result measured by scanning tunneling spectroscopy (STS) on a single S-vacancy defect.[91] Energy levels related to S-vacancies have also been observed in the same energy range below CB by deep level transient spectroscopy.[94]

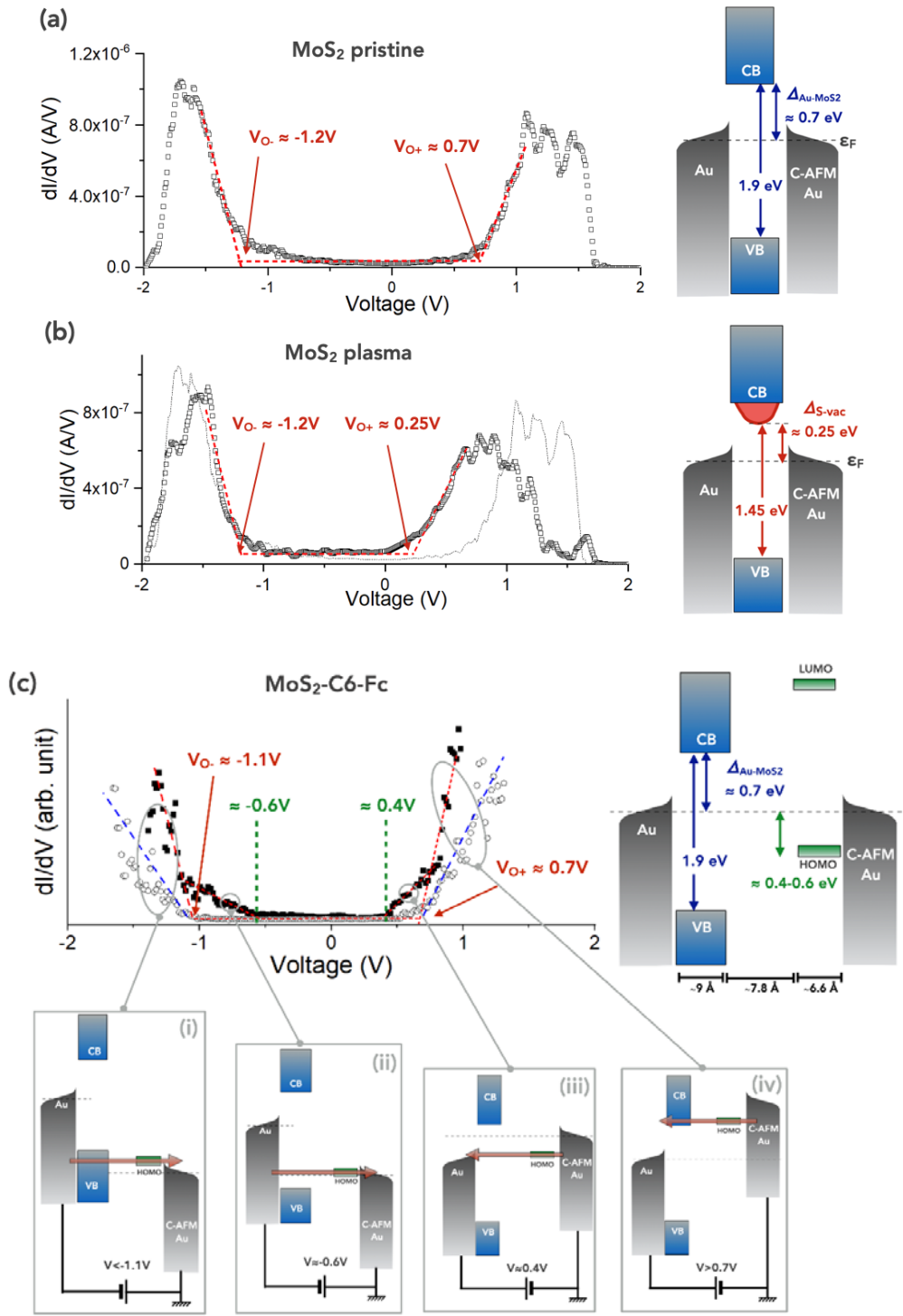

**Figure 9**. First derivative $\partial I/\partial V$-$V$ of $I$-$V$ spectra and proposed energy diagrams at zero bias of (a) pristine MoS2, (b) plasma-treated MoS2 and (c) ferrocene-grafted MoS2 samples. In the panel (b), the thin line is the $\partial I/\partial V$-$V$ of the pristine MoS2 shown in the panel (a) and redrawn for comparison. In panels (a) and (b), the $\partial I/\partial V$-$V$ curves were numerically calculated from the mean $\bar{I}$-$V$ curves shown in Figure 8a and 8b (see the Experimental Section). In the panel (c), the $\partial I/\partial V$-

$V$ curves were directly measured by a lock-in technique. The bottom gray-line frames ((*i*) to (*iv*)) schematically depict the proposed electron transfer mechanisms at several voltages (see text).

The results for the ferrocene-grafted MoS$_2$ junctions are more delicate to interpret. From the $\partial I/\partial V$-$V$ spectrum which was numerically calculated from the mean $\bar{I}$-$V$ curve shown in Figure 8c, we recover the same onset voltages as for the pristine MoS$_2$ ($V_{o+} \approx 0.7$ V and $V_{o-} \approx -1.2$ V, Figure S12). This feature indicates a good healing of the S-vacancy defects by the thiol-terminated molecules, in agreement with our Raman spectroscopy and PL measurements (*vide supra*), and as also indirectly concluded from the reported characteristic modification of the molecularly functionalized MoS$_2$ transistor parameters.[28,46] However, no conclusions can be drawn about the energy position of the molecular orbitals of the ferrocenyl moiety. From data reported in the literature on metal/alkyl-ferrocene/metal molecular junctions, the highest occupied molecular orbital (HOMO) of ferrocene is expected to be at ≈ 0.3-0.4 eV below the Au Fermi energy (HOMO of ferrocene at -5.1 eV from the vacuum level and the Au work function at -4.7/-4.8 eV).[95,96] For a ferrocene chemically grafted to MoS$_2$ (via a short alkyl chain (CH$_2$)$_2$), DFT calculations predict the HOMO at ~0.6 eV above the VB of MoS$_2$.[34] Thus, we can expect to detect a contribution of the HOMO of ferrocene to the electron transfer properties if it is energetically located in the band gap of MoS$_2$. The lowest unoccupied molecular orbital (LUMO) of ferrocene is higher (≈-0.5 V vs. vacuum) and does not contribute to the electron transport. To have a better sensitivity and energy resolution, we have also directly measured $\partial I/\partial V$ by a lock-in technique (see the Experimental Section). Out of 158 $\partial I/\partial V$-$V$ curves measured on the ferrocene-grafted MoS$_2$ junctions, two typical cases can be distinguished as shown in Figure 9c. A majority of curves gives the same behavior as the pristine MoS$_2$, depicted as white circles in Figure 9c, with almost similar onset voltages: $V_{o+} \approx 0.7$ V and $V_{o-} \approx -1.1$ V. For about 12% of the dataset, the $\partial I/\partial V$-$V$ curves show a marked difference exhibiting two bumps at lower onset voltages of ≈ 0.4 V and ≈ -0.6 V. These bumps are attributed to electron transport through the HOMO of ferrocene as depicted by the light gray framed energy diagrams in Figure 9c. Now, we still assume an energy pinning at the Au electrode/MoS$_2$ and that the applied voltage mainly drops through the alkyl chains linking MoS$_2$ to ferrocene. Additionally, a slight voltage-induced shift of the HOMO of ferrocene with respect to the Fermi energy of C-AFM Au can be also foreseen. At $V \approx 0.4$ V (frame (*iii*) in Figure 9c), the HOMO of ferrocene is entering the energy windows and the electrons are transported resonantly through the HOMO and then by tunneling through the band gap of MoS$_2$. At higher voltages, the electrons are also transported via the MoS$_2$ CB and the HOMO of ferrocene (frame (*iv*)). The same behavior holds at negative voltages for the HOMO and VB of MoS$_2$ (frames (*i*) and (*ii*)). Thus, we assume that, for this small fraction of the dataset, we have detected the HOMO of the neutral Fc lying at *ca.* 0.4-0.6 eV below the electrode Fermi energy, in agreement with our DFT calculations (*vide infra*). For the majority of the dataset, we do not detect any molecular orbital in the MoS$_2$ band gap, consistent with the observation that a majority of the ferrocene moieties are in their oxidized state (as also inferred from the XPS data, Fig. 3i), for which the level is deeper in energy and masked by the MoS$_2$ VB. At a macroscopic scale, for ferrocene-grafted MoS$_2$, the features of the MoS$_2$ valence band determined by UPS are not distinguishable due to the presence of ferrocene molecules on the MoS$_2$ surface and we cannot precisely position the VBM. Nevertheless, the absence of the HOMO peak below the Fermi energy can be rationalized by the fact that, at this macroscopic scale, the dominant ferrocenium moieties (mixed with MoS$_2$ VB) contribute mainly to the UPS spectrum, in agreement with the above discussion of the $\partial I/\partial V$-$V$ curves.

### 3.5. Density Functional Theory (DFT) calculations.

In order to shed further light on the electronic structure of the MoS$_2$/Au systems, first-principles periodic calculations based on DFT were also carried out. For the sake of simplicity in our models, we assume that the pristine and ferrocene-grafted MoS$_2$ systems do not exhibit any sulfur defect while the defective MoS₂ system exhibits one sulfur vacancy per supercell. To create the MoS₂-Au interface, the Au (111) surface was modeled by a slab of 5 layers strained by only 0.18% to match the lattice constant of MoS₂ (see Figure 10a). The grafted MoS₂ is created by grafting the 6-(ferrocenyl)hexanethiol molecule on the defective MoS₂ so that the sulfur of the hexanethiol chain replaces the removed MoS₂ sulfur atom. Once the three MoS₂ models are created, periodic DFT calculations are performed by using the QuantumATK/2022.03 package.[97,98]

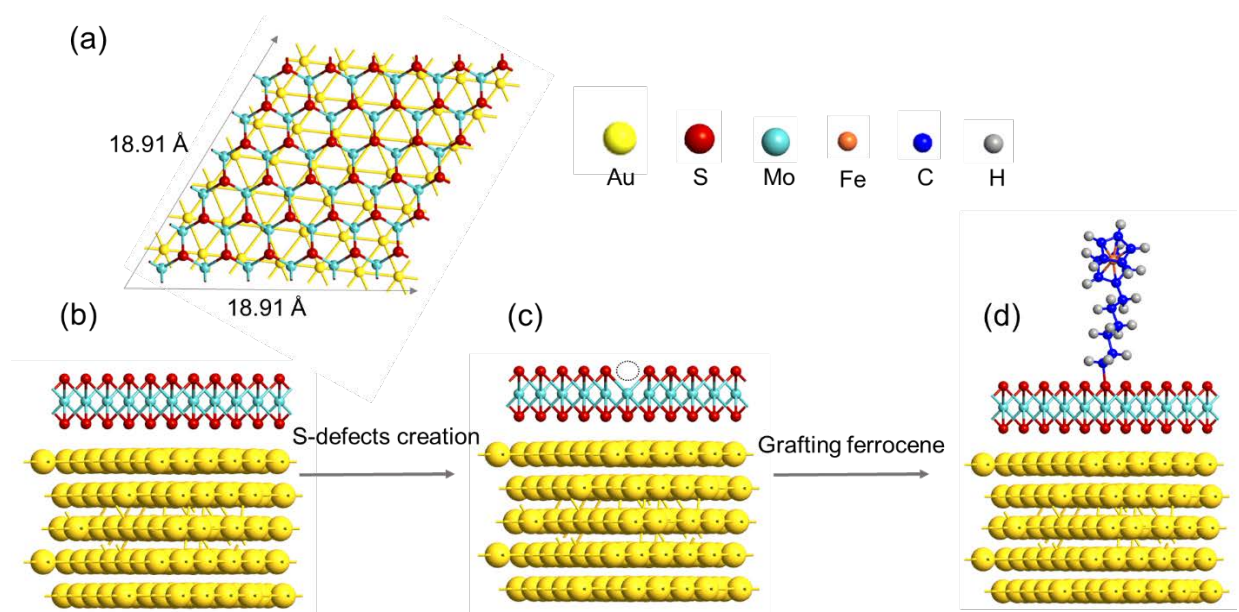

**Figure 10.** (a) Top view of the simulated supercell of the MoS₂/Au interface with the Au (111) surface strained by 0.18%. A side view of the optimized (b) pristine MoS₂/Au, (c) defective MoS₂/Au created by removing one sulfur atom from pristine MoS₂/Au, and (d) functionalized MoS₂/Au created by grafting 6-(ferrocenyl)hexanethiol on defective MoS₂/Au.

First, the atomic positions of the three MoS₂ structures (pristine MoS₂/Au, defective MoS₂/Au and ferrocene-grafted MoS₂/Au) were fully optimized until the final forces acting on the atoms are less than 0.02 eV/Å. A vacuum region (~22 Å) normal to the surface is added to minimize the interaction between adjacent slabs. We use the Perdew-Burke-Ernzerhof (PBE) functional within the generalized gradient approximation (GGA)[99] for geometric optimization whereas the hybrid functional HSE06[100,101] is used to calculate the electronic structure. This hybrid functional, with a screening parameter $\omega = 0.11$ bohr$^{-1}$ and an exact exchange parameter $a = 0.25$, represents a reasonable compromise between computational cost and quality of the results.[102] The valence electrons are described within the LCAO approximation with single zeta plus polarization (double-zeta plus polarization basis set) for gold atoms (other atoms) whilst the core electrons are described by the norm-conserving Troullier-Martins pseudopotentials.[103] We used a density mesh cutoff of 100 Ha and a (3×3×1) Monkhorst-Pack k-point sampling for atomic optimization and (12×12×1) for electronic structure. Note that spin-orbit coupling (SOC) effect has not been included for these DFT calculations due to the large size of the simulated MoS$_2$/Au interface.[104,105,106]

Within this framework, we compute a work function of 4.04 eV for pristine MoS$_2$/Au, a lower value compared to experimental values from literature (4.54 eV-4.9 eV).[107,108,109] This result is fully expected due to the use of short-range localized atomic orbitals that induces an artificial push-back effect at the interface and thus reduces the metal work function. To solve this numerical problem, we added a layer of ghost atoms in the vacuum next to the MoS$_2$ surface to better describe the tail of the electron density, an approach already implemented in QuantumATK/2022.03 package.[97] The obtained results indicate that the use of three layers of gold ghost atoms at a distance of 3 Å from the bottom sulfur layer of MoS$_2$ (Figure S14) provides a work function of 4.75 eV, which is in good agreement with the reported experimental values. After ferrocene grafting on defective MoS$_2$, the work function is decreased by 0.53 eV, which is consistent with *n*-type doping by the grafting molecule.

For detailed analysis of the electronic structures of the MoS$_2$/Au systems, we examine the partial density of states (PDOS) as shown in Figure 11. The results reveal that the conduction band minimum (CBM) and valence band maximum (VBM) of pristine MoS$_2$ are respectively located at 1.14 eV and -1.33 eV with respect to the Fermi level. The calculated value of VBM is in excellent agreement with the experimental value of 1.34 eV measured by UPS (Fig. S13) and *ca.* 1.2 eV by C-AFM (Fig. 9). Although the calculated band gap of pristine $MoS_2$ by using HSE06 functional is larger than the experimental value (2.47 eV vs 1.9 eV), the experimental energy alignment of the three investigated systems is well described. The difference can be attributed to the gold surface polarization (image charge effects), which is not captured by approximated exchange-correlation functionals.[110] In fact, we clearly notice the creation of S-vacancy states in the energy range of 0.35 eV above the Fermi level for defective $MoS_2$ (Figure 11b), which is in good agreement with C-AFM experimental measurements revealing the reduction of $V_{O+}$ due to the creation of S-vacancy defect state at 0.25 eV above the Fermi level (Fig. 9b). Moreover, the PDOS analysis shows that this S-vacancy defect state disappears after ferrocene grafting on defective $MoS_2$, which indicates a good healing of the S-vacancy defects (Figure 11c). Noticeably, a new molecular state appears in the band gap of $MoS_2$, at 0.45 eV below the Femi level that is attributed to the HOMO level of ferrocene molecule (Fe d-orbitals and C p-orbitals, Figure 11c). This feature agrees with the experimental $\partial I/\partial V$-$V$ curves datasets suggesting electron transport through HOMO of ferrocene due to the existence of two bumps at lower onset voltage of 0.4 and 0.6 V (Fig. 9c).

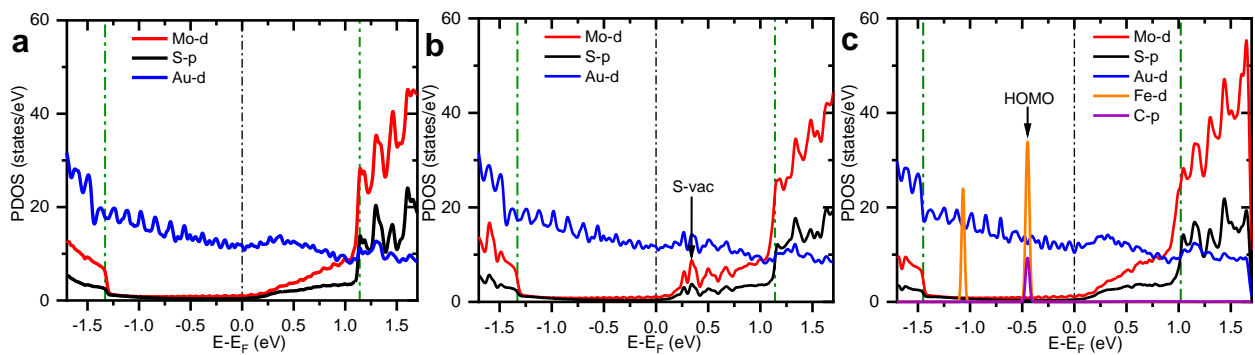

**Figure 11.** PDOS for (a) pristine $MoS_2$/Au, (b) defective $MoS_2$/Au and (c) ferrocene-grafted $MoS_2$/Au. Black dashed line refers to the Fermi level and green dashed lines indicate valence band maximum (VBM) and conduction band minimum (CBM) of $MoS_2$.

## 4. Conclusions

In this work, the reaction of thiol-substituted ferrocenyl derivative on defective monolayered $MoS_2$ resulted in the grafting of $3.1 \times 10^{12}$ molecules per $cm^2$ (i.e. >90% of the defect sites) which is consistent with the formation of a densely packed redox-active monolayer. Furthermore, the redox-active properties of the grafted ferrocene were not altered after this immobilization step, as supported by cyclic voltammetry. A fine and in-depth understanding of the electronic properties of electroactive ferrocene-functionalized $MoS_2$ has been gained through the combination of multiple experimental techniques and theoretical modeling. The out-of-plane electron transport measurements by conductive AFM has enabled the energy diagram of different $MoS_2$/Au junctions at each step (pristine, defective and molecularly functionalized $MoS_2$) to be built. As an effect of molecular functionalization, two characteristic bumps were observed in the $\partial I/\partial V - V$ traces of the ferrocene-functionalized $MoS_2$/Au junction which were attributed to the electron transport through the ferrocene HOMO located in the $MoS_2$ band gap. Importantly, the electronic structures of the different $MoS_2$/Au systems determined from conductive AFM measurements were in good agreement with the partial density of states (PDOS) extracted from DFT theoretical calculations.

The knowledge acquired for ferrocenyl functionalized $MoS_2$ represents an essential step towards the implementation of novel electrochemically switchable devices. Toward this goal, it is anticipated that changing the redox state of the grafted molecule by the applied voltage will influence both spin-orbit coupling and the spin diffusion length of functionalized $MoS_2$ and thus the measurable tunnel magnetoresistance output.

## ASSOCIATED CONTENT

**Supporting Information**. Details on the preparation and characterization of the $MoS_2$ samples, additional Raman spectra and fitting, AFM images, cyclic voltammetry data, effective electron mobility measurements, optical images of the samples for the in-plane electron transfer measurements, estimation of the contact area for the electronic transport measurements, complementary electronic transport data, UPS spectra, and $MoS_2$/Au structures with ghost atoms (DFT calculations). The Supporting Information is available free of charge on the ACS Publications website.

## AUTHOR INFORMATION

**Corresponding Author**

* **Bruno Fabre** ― CNRS, ISCR (Institut des Sciences Chimiques de Rennes) - UMR6226, Univ Rennes, Rennes F-35000, France. E-mail: bruno.fabre@univ-rennes1.fr

**Author Contributions**

All authors participated in conceiving the project, designing the experiments, analyzing the data and writing the manuscript. T. N. N. L. and J. C. L. carried out the exfoliation of $MoS_2$, performed and analyzed the Raman and XPS spectroscopy, photoluminescence, AFM and electrochemical data. K. K. and D. V. performed and analyzed the in-plane and out-of-plane electronic transport data. I. A., C. V. D. and J. C. performed the theoretical calculations. All authors contributed to the results discussion. All authors participated in editing the paper and have given approval to the final version of the manuscript.

**Funding Sources**

This project has received financial support from ANR (Agence Nationale de la Recherche).

ACKNOWLEDGMENT

The authors gratefully acknowledge ANR (Agence Nationale de la Recherche) for financial support of the ECOTRAM program (ECOTRAM, grant number ANR-20-CE09-0018-01). T. N. N. L. is grateful to ANR for the funding of his Ph.D. The XPS measurements have been performed on the ASPHERYX platform (ScanMAT, UAR 2025 University of Rennes-CNRS; CPER-FEDER 2015-2020). S. Ababou-Girard (Institut de Physique de Rennes, UMR6251, France) is fully acknowledged for some preliminary XPS measurements. The computational resources were provided by the Consortium des "Equipements de Calcul Intensif" (CÉCI) funded by the Belgian National Fund for Scientific Research (F.R.S.-FNRS) under Grant 2.5020.11. J.C. is an FNRS Research Director.

REFERENCES

(1) Rao, C. N. R.; Gopalakrishnan, K.; Maitra, U. Comparative Study of Potential Applications of Graphene, MoS$_2$, and Other Two-Dimensional Materials in Energy Devices, Sensors, and Related Areas. *ACS Appl. Mater. Interfaces* **2015**, *7*, 7809-7832.

(2) Yin, X.; Tang, C. S.; Zheng, Y.; Gao, J.; Wu, J.; Zhang, H.; Chhowalla, M.; Chen, W.; Wee, A. T. S. Recent Developments in 2D Transition Metal Dichalcogenides: Phase Transition and Applications of the (Quasi-)Metallic Phases. *Chem. Soc. Rev.* **2021**, *50*, 10087-10115.

(3) Wang, Q. H.; Kalantar-Zadeh, K.; Kis, A.; Coleman, J. N.; Strano, M. S. Electronics and Optoelectronics of Two-Dimensional Transition Metal Dichalcogenides. *Nature Nanotech.* **2012**, *7*, 699-712.

(4) Choudhuri, I.; Bhauriyal, P.; Pathak, B. Recent Advances in Graphene-like 2D Materials for Spintronics Applications. *Chem. Mater.* **2019**, *31*, 8260-8285.

(5) Chhowalla, M.; Shin, H. S.; Eda, G.; Li, L.-J.; Loh, K. P.; Zhang, H. The Chemistry of Two-Dimensional Layered Transition Metal Dichalcogenide Nanosheets. *Nature Chem.* **2013**, *5*, 263-275.

(6) Chia, X.; Pumera, M. Layered Transition Metal Dichalcogenide Electrochemistry: Journey across the Periodic Table. *Chem. Soc. Rev.* **2018**, *47*, 5602-5613.

(7) Voiry, D.; Yang, J.; Chhowalla, M. Recent Strategies for Improving the Catalytic Activity of 2D TMD Nanosheets Toward the Hydrogen Evolution Reaction. *Adv. Mater.* **2016**, *28*, 6197-6206.

(8) Zhu, C. (R.); Gao, D.; Ding, J.; Chao, D.; Wang, J. TMD-Based Highly Efficient Electrocatalysts Developed by Combined Computational and Experimental Approaches. *Chem. Soc. Rev.* **2018**, *47*, 4332-4356.

(9) Li, X.; Tao, L.; Chen, Z.; Fang, H.; Li, X.; Wang, X.; Xu, J.-B.; Zhu, H. Graphene and Related Two-Dimensional Materials: Structure-Property Relationships for Electronics and Optoelectronics. *Appl. Phys. Rev.* **2017**, *4*, 021306.

(10) Mak, K. F.; Lee, C.; Hone, J.; Shan, J.; Heinz, T. F. Atomically Thin $MoS_2$: a New Direct-Gap Semiconductor. *Phys. Rev. Lett.* **2010**, *105*, 136805.

(11) Schaibley, J. R.; Yu, H.; Clark, G.; Rivera, P.; Ross, J. S.; Seyler, K. L.; Yao, W.; Xu, X. Valleytronics in 2D Materials. *Nat. Rev. Mater.* **2016**, *1*, 16055.

(12) Ye, Y.; Xiao, J.; Wang, H.; Ye, Z.; Zhu, H.; Zhao, M.; Wang, Y.; Zhao, J.; Yin, X.; Zhang, X. Electrical Generation and Control of the Valley Carriers in a Monolayer Transition Metal Dichalcogenide. *Nature Nanotech.* **2016**, *11*, 598-602.

(13) Dankert, A.; Pashaei, P.; Kamalakar, M. V.; Gaur, A. P. S.; Sahoo, S.; Rungger, I.; Narayan, A.; Dolui, K.; Hoque, Md. A.; Patel, R. S. et al. Spin-Polarized Tunneling through Chemical Vapor Deposited Multilayer Molybdenum Disulfide. *ACS Nano* **2017**, *11*, 6389-6395.

(14) Yan, W.; Txoperena, O.; Llopis, R.; Dery, H.; Hueso, L. E.; Casanova, F. A Two-Dimensional Spin Field-Effect Switch. *Nat. Commun.* **2016**, *7*, 13372.

(15) Bertolazzi, S.; Gobbi, M.; Zhao, Y.; Backes, C.; Samori, P. Molecular Chemistry Approaches for Tuning the Properties of Two-Dimensional Transition Metal Dichalcogenides. *Chem. Soc. Rev.* **2018**, *47*, 6845-6888.

(16) Stergiou, A.; Tagmatarchis, N. Molecular Functionalization of Two-Dimensional $MoS_2$ Nanosheets. *Chem. Eur. J.* **2018**, *24*, 18246-18257.

(17) Chen, X.; McDonald, A. R. Functionalization of Two-Dimensional Transition-Metal Dichalcogenides. *Adv. Mater.* **2016**, *28*, 5738-5746.

(18) Zhao, Y.; Manoj Gali, S.; Wang, C.; Pershin, A.; Slassi, A.; Beljonne, D.; Samori, P. Molecular Functionalization of Chemically Active Defects in $WSe_2$ for Enhanced Opto-Electronics. *Adv. Funct. Mater.* **2020**, *30*, 2005045.

(19) Kumari, S.; Chouhan, A.; Sharma, O. P.; Tawfik, S. A.; Tran, K.; Spencer, M. J. S.; Bhargava, S. K.; Walia, S.; Ray, A.; Khatri, O. P. Surface Functionalization of $WS_2$ Nanosheets with Alkyl Chains for Enhancement of Dispersion Stability and Tribological Properties. *ACS Appl. Mater. Interfaces* **2022**, *14*, 1334-1346.

(20) Voiry, D.; Goswami, A.; Kappera, R.; de Carvalho Castro e Silva, C.; Kaplan, D.; Fujita, T.; Chen, M.; Asefa, T.; Chhowalla, M. Covalent Functionalization of Monolayered Transition Metal Dichalcogenides by Phase Engineering. *Nature Chem.* **2015**, *7*, 45-49.

(21) Vera-Hidalgo, M.; Giovanelli, E.; Navio, C.; Perez, E. M. Mild Covalent Functionalization of Transition Metal Dichalcogenides with Maleimides: A "Click" Reaction for 2H-$MoS_2$ and $WS_2$. *J. Am. Chem. Soc.* **2019**, *141*, 3767-3771.

(22) Tuci, G.; Mosconi, D.; Rossin, A.; Luconi, L.; Agnoli, S.; Righetto, M.; Pham-Huu, C.; Ba, H.; Cicchi, S.; Granozzi, G. et al. Surface Engineering of Chemically Exfoliated $MoS_2$ in a "Click": How To Generate Versatile Multifunctional Transition Metal Dichalcogenides-Based Platforms. *Chem. Mater.* **2018**, *30*, 8257-8269.

(23) Chu, X. S.; Yousaf, A.; Li, D. O.; Tang, A. A.; Debnath, A.; Ma, D.; Green, A. A.; Santos, E. J. G.; Wang, Q. H. Direct Covalent Chemical Functionalization of Unmodified Two-Dimensional Molybdenum Disulfide. *Chem. Mater.* **2018**, *30*, 2112-2128.

(24) Lihter, M.; Graf, M.; Iveković, D.; Zhang, M.; Shen, T.-H.; Zhao, Y.; Macha, M.; Tileli, V.; Radenovic, A. Electrochemical Functionalization of Selectively Addressed $MoS_2$ Nanoribbons for Sensor Device Fabrication. *ACS Appl. Nano Mater.* **2021**, *4*, 1076-1084.

(25) Yan, E. X.; Cabán-Acevedo, M.; Papadantonakis, K. M.; Brunschwig, B. S.; Lewis, N. S. Reductant-Activated, High-Coverage, Covalent Functionalization of 1T′-$MoS_2$. *ACS Materials Lett.* **2020**, *2*, 133-139.

(26) Cho, K.; Pak, J.; Chung, S.; Lee, T. Recent Advances in Interface Engineering of Transition-Metal Dichalcogenides with Organic Molecules and Polymers. *ACS Nano* **2019**, *13*, 9713-9734.

(27) Chen, X.; Berner, N. C.; Backes, C.; Duesberg, G. S.; McDonald, A. R. Functionalization of Two-Dimensional MoS$_2$: On the Reaction Between MoS$_2$ and Organic Thiols. *Angew. Chem. Int. Ed.* **2016**, *55*, 5803-5808.

(28) Cho, K.; Min, M.; Kim, T.-Y.; Jeong, H.; Pak, J.; Kim, J.-K.; Jang, J.; Yun, S. J.; Lee, Y. H.; Hong, W.-K. et al. T. Electrical and Optical Characterization of MoS$_2$ with Sulfur Vacancy Passivation by Treatment with Alkanethiol Molecules. *ACS Nano* **2015**, *9*, 8044-8053.

(29) Ippolito, S.; Kelly, A. G.; Furlan de Oliveira, R.; Stoeckel, M.-A.; Iglesias, D.; Roy, A.; Downing, C.; Bian, Z.; Lombardi, L.; Abdul Samad, Y. et al. Covalently Interconnected Transition Metal Dichalcogenide Networks via Defect Engineering for High-Performance Electronic Devices. *Nature Nanotech.* **2021**, *16*, 592-598.

(30) Zhao, Y.; Gobbi, M.; Hueso, L. E.; Samori, P. Molecular Approach to Engineer Two-Dimensional Devices for CMOS and beyond-CMOS Applications. *Chem. Rev.* **2022**, *122*, 50-131.

(31) Ping, J.; Fan, Z.; Sindoro, M.; Ying, Y.; Zhang, H. Recent Advances in Sensing Applications of Two-Dimensional Transition Metal Dichalcogenide Nanosheets and Their Composites. *Adv. Funct. Mater.* **2017**, *27*, 1605817.

(32) Chou S. S.; De, M.; Kim, J.; Byun, S.; Dykstra, C.; Yu, J.; Huang, J.; Dravid, V. P. Ligand Conjugation of Chemically Exfoliated MoS$_2$. *J. Am. Chem. Soc.* **2013**, *135*, 4584-4587.

(33) Zhao, Y.; Bertolazzi, S.; Maglione, M. S.; Rovira, C.; Mas-Torrent, M.; Samori, P. Molecular Approach to Electrochemically Switchable Monolayer MoS$_2$ Transistors. *Adv. Mater.* **2020**, *32*, 2000740.

(34) Li, D..; Zhang, G.; Hu, Y. Y.; Shang, Y. Electronic and Transport Properties of Covalent Functionalized Monolayer MoS$_2$ by Ferrocene Derivatives. *JOM* **2023**, *75*, 603-613.

(35) Wang, Y.; Kim, C.-H.; Yoo, Y.; Johns, J. E.; Frisbie, C. D. Field Effect Modulation of Heterogeneous Charge Transfer Kinetics at Back-Gated Two-Dimensional $MoS_2$ Electrodes. *Nano Lett.* **2017**, *17*, 7586-7592.

(36) Kuo, D.-Y.; Rice, P. S.; Raugei, S.; Cossairt, B. M. Charge Transfer in Metallocene Intercalated Transition Metal Dichalcogenides. *J. Phys. Chem. C* **2022**, *126*, 13994-14002.

(37) Du, H.-Y.; Huang, Y.-F.; Wong, D.; Tseng, M.-F.; Lee, Y.-H.; Wang, C.-H.; Lin, C.-L.; Hoffman, G.; Chen, K.-H.; Chen, L.-C. Nanoscale Redox Mapping at the $MoS_2$-Liquid Interface. *Nat. Commun.* **2021**, 12:1321.

(38) Liu, F.; Wu, W.; Bai, Y.; Chae, S.H.; Li, Q.; Wang, J.; Hone, J.; Zhu, X.-Y. Disassembling 2D van der Waals Crystals into Macroscopic Monolayers and Reassembling into Artificial Lattices. *Science* **2020**, *367*, 903.

(39) Zhan, Y.; Liu, Z.; Najmaei, S.; Ajayan, P. M.; Lou, J. Large-Area Vapor-Phase Growth and Characterization of $MoS_2$ Atomic Layers on a $SiO_2$ Substrate. *Small* **2012**, *8*, 966–971.

(40) Li, H.; Zhang, Q.; Yap, C. C. R.; Tay, B. K.; Edwin, T. H. T.; Olivier, A.; Baillargeat, D. From Bulk to Monolayer $MoS_2$: Evolution of Raman Scattering. *Adv. Funct. Mater.* **2012**, *22*, 1385–1390.

(41) Mignuzzi, S.; Pollard, A. J.; Bonini, N.; Brennan, B.; Gilmore, I. S.; Pimenta, M. A.; Richards, D.; Roy, D. Effect of Disorder on Raman Scattering of Single-Layer $MoS_2$. *Phys. Rev. B - Condens. Matter Mater. Phys.* **2015**, *91*, 1–7.

(42) Lin, Z.; Carvalho, B. R.; Kahn, E.; Lv, R.; Rao, R.; Terrones, H.; Pimenta, M. A.; Terrones, M. Defect Engineering of Two-Dimensional Transition Metal Dichalcogenides. *2D Mater.* **2016**, *3*, 022002.

(43) Aryeetey, F.; Ignatova, T.; Aravamudhan, S. Quantification of Defects Engineered in Single Layer $MoS_2$. *RSC Adv.* **2020**, *10*, 22996–23001.

(44) Wong, C. P. Y.; Koek, T. J. H.; Liu, Y.; Loh, K. P.; Goh, K. E. J.; Troadec, C.; Nijhuis, C. A. Electronically Transparent Graphene Barriers Against Unwanted Doping of Silicon. *ACS Appl. Mater. Interfaces* **2014**, *6*, 20464–20472.

(45) Nguyen, E. P.; Carey, B. J.; Ou, J. Z.; Van Embden, J.; Gaspera, E. D.; Chrimes, A. F.; Spencer, M. J. S.; Zhuiykov, S.; Kalantar-Zadeh, K.; Daeneke, T. Electronic Tuning of 2D $MoS_2$ through Surface Functionalization. *Adv. Mater.* **2015**, *27*, 6225–6229.

(46) Bertolazzi, S.; Bonacchi, S.; Nan, G.; Pershin, A.; Beljonne, D.; Samorì, P. Engineering Chemically Active Defects in Monolayer $MoS_2$ Transistors via Ion-Beam Irradiation and Their Healing via Vapor Deposition of Alkanethiols. *Adv. Mater.* **2017**, *29*, 1606760.

(47) Cançado, L. G.; Jorio, A.; Ferreira, E. H. M.; Stavale, F.; Achete, C. A.; Capaz, R. B.; Moutinho, M. V. O.; Lombardo, A.; Kulmala, T. S.; Ferrari, A. C. Quantifying Defects in Graphene via Raman Spectroscopy at Different Excitation Energies. *Nano Lett* **2011**, *11*, 3190–3196.

(48) Lu, J.; Carvalho, A.; Chan, X. K.; Liu, H.; Liu, B.; Tok, E. S.; Loh, K. P.; Castro Neto, A. H.; Sow, C. H. Atomic Healing of Defects in Transition Metal Dichalcogenides. *Nano Lett.* **2015**, *15*, 3524–3532.

(49) Ma, D.; Wang, Q.; Li, T.; He, C.; Ma, B.; Tang, Y.; Lu, Z.; Yang, Z. Repairing Sulfur Vacancies in the $MoS_2$ Monolayer by using CO, NO and $NO_2$ Molecules. *J. Mater. Chem. C* **2016**, *4*, 7093–7101.

(50) Plechinger, G.; Castellanos-gomez, A.; Buscema, M.; van der Zant, H. S. J.; Steele, G. A.; Kuc, A.; Heine, T.; Schüller, C.; Korn T. Control of Biaxial Strain in Single-Layer Molybdenite using Local Thermal Expansion of the Substrate. *2D Mater.* **2015**, *2*, 15006.

(51) Lloyd, D.; Liu, X.; Christopher, J. W.; Cantley, L.; Wadehra, A.; Kim, B. L.; Goldberg, B. B.; Swan, A. K.; Bunch, J. S. Band Gap Engineering with Ultralarge Biaxial Strains in Suspended Monolayer $MoS_2$. *Nano Lett.* **2016**, *16*, 5836–5841.

(52) Conley, H. J.; Wang, B.; Ziegler, J. I.; Haglund, R. F.; Pantelides, S. T.; Bolotin, K. I. Bandgap Engineering of Strained Monolayer and Bilayer MoS$_2$. *Nano Lett.* **2013**, *13*, 3626–3630.

(53) Mouri, S.; Miyauchi, Y.; Matsuda, K. Tunable Photoluminescence of Monolayer MoS$_2$ via Chemical Doping. *Nano Lett.* **2013**, *13*, 5944-5948.

(54) Wang, Y.; Slassi, A.; Stoeckel, M.-A.; Bertolazzi, S.; Cornil, J.; Beljonne, D.; Samorì, P. Doping of Monolayer Transition-Metal Dichalcogenides via Physisorption of Aromatic Solvent Molecules. *J. Phys. Chem. Lett.* **2019**, *10*, 540–547.

(55) Mak, K. F.; He, K.; Lee, C.; Lee, G. H.; Hone, J.; Heinz, T. F.; Shan, J. Tightly Bound Trions in Monolayer MoS$_2$. *Nat. Mater.* **2013**, *12*, 207–211.

(56) Zhang, X.; Nan, H.; Xiao, S.; Wan, X.; Ni, Z.; Gu, X.; Ostrikov, K. Shape-Uniform, High-Quality Monolayered MoS$_2$ Crystals for Gate-Tunable Photoluminescence. *ACS Appl. Mater. Interfaces* **2017**, *9*, 42121-42130.

(57) Tongay, S.; Zhou, J.; Ataca, C.; Liu, J.; Kang, J. S.; Matthews, T. S.; You, L.; Li, J.; Grossman, J. C.; Wu, J. Broad-Range Modulation of Light Emission in Two-Dimensional Semiconductors by Molecular Physisorption Gating. *Nano Lett.* **2013**, *13*, 2831–2836.

(58) Nan, H.; Wang, Z.; Wang, W.; Liang, Z.; Lu, Y.; Chen, Q.; He, D.; Tan, P.; Miao, F.; Wang, X. et al. Strong Photoluminescence Enhancement of MoS$_2$ through Defect Engineering and Oxygen Bonding. *ACS Nano* **2014**, *8*, 5738–5745.

(59) Grünleitner, T.; Henning, A.; Bissolo, M.; Zengerle, M.; Gregoratti, L.; Amati, M.; Zeller, P.; Eichhorn, J.; Stier, A. V.; Holleitner, A. W. et al. Real-Time Investigation of Sulfur Vacancy Generation and Passivation in Monolayer Molybdenum Disulfide via in situ X-ray Photoelectron Spectromicroscopy. *ACS Nano* **2022**, *16*, 20364-20375.

(60) Gao, L.; Liao, Q.; Zhang, X.; Liu, X.; Gu, L.; Liu, B.; Du, J.; Ou, Y.; Xiao, J.; Kang, Z. et al. Defect-Engineered Atomically Thin MoS$_2$ Homogeneous Electronics for Logic Inverters. *Adv Mater* **2020**, *32*, 1906646.

(61) Ganta, D.; Sinha, S.; Haasch, R. T. 2-D Material Molybdenum Disulfide Analyzed by XPS. *Surf. Sci. Spectra* **2014**, *21*, 19-27.

(62) Li, L.; Qin, Z.; Ries, L.; Hong, S.; Michel, T.; Yang, J.; Salameh, C.; Bechelany, M.; Miele, P.; Kaplan, D. et al. Role of Sulfur Vacancies and Undercoordinated Mo Regions in $MoS_2$ Nanosheets toward the Evolution of Hydrogen. *ACS Nano* **2019**, *13*, 6824−6834.

(63) Taylor, A. W.; Licence, P. X-Ray Photoelectron Spectroscopy of Ferrocenyl- and Ferrocenium-Based Ionic Liquids. *ChemPhysChem* **2012**, *13*, 1917-1926.

(64) Ali, G. A. M.; Megiel, E.; Cieciorski, P.; Thalji, M. R.; Romanski, J.; Algarni, H.; Chong, K. F. Ferrocene Functionalized Multi-Walled Carbon Nanotubes as Supercapacitor Electrodes. *J. Mol. Liq.* **2020**, *318*, 114064.

(65) Graat, P. C. J.; Somers, M. A. J. Simultaneous Determination of Composition and Thickness of Thin Iron-Oxide Films from XPS Fe 2p Spectra. *Appl. Surf. Sci.* **1996**, *100-101*, 36-40.

(66) Splendiani, A.; Sun, L.; Zhang, Y.; Li, T.; Kim, J.; Chim, C.-Y.; Galli, G.; Wang, F. Emerging Photoluminescence in Monolayer $MoS_2$. *Nano Lett.* **2010**, *10*, 1271-1275.

(67) Li, H.; Wu, J.; Yin, Z.; Zhang, H. Preparation and Applications of Mechanically Exfoliated Single-Layer and Multilayer $MoS_2$ and $WSe_2$ Nanosheets. *Acc. Chem. Res.* **2014**, *47*, 1067-1075.

(68) Smalley, J. F.; Finklea, H. O.; Chidsey, C. E. D.; Linford, M. R.; Creager, S. E.; Ferraris, J. P.; Chalfant, K.; Zawodzinsk, T.; Feldberg, S. W.; Newton, M. D. Heterogeneous Electron-Transfer Kinetics for Ruthenium and Ferrocene Redox Moieties through Alkanethiol Monolayers on Gold. *J. Am. Chem. Soc.* **2003**, *125*, 2004-2013.

(69) Chidsey, C. E. D.; Bertozzi, C. R.; Putvinski, T. M.; Mujsce, A. M. Coadsorption of Ferrocene-Terminated and Unsubstituted Alkanethiols on Gold: Electroactive Self-Assembled Monolayers. *J. Am. Chem. Soc.* **1990**, *112*, 4301-4306.

(70) Bard, A. J.; Faulkner, L. R. Electrochemical Methods. Fundamentals and Applications, Wiley: New York, 1980, p. 522.

(71) Radisavljevic, B.; Radenovic, A.; Brivio, J.; Giacometti, V.; Kis, A. Single-Layer MoS$_2$ Transistors. *Nat. Nanotechnol.* **2011**, *6*, 147-150.

(72) Radisavljevic, B.; Kis, A. Mobility Engineering and a Metal-Insulator Transition in Monolayer MoS$_2$. *Nat. Mater.* **2013**, *12*, 815-820.

(73) Novoselov, K. S.; Jiang, D.; Schedin, F.; Booth, T. J.; Khotkevich, V. V.; Morozov, S. V.; Geim, A. K. Two-Dimensional Atomic Crystals. *Proc. Natl. Acad. Sci.* **2005**, *102*, 10451-10453.

(74) Di Bartolomeo, A.; Grillo, A.; Urban, F.; Iemmo, L.; Giubileo, F.; Luongo, G.; Amato, G.; Croin, L.; Sun, L.; Liang, S.-J.; Ang, L. K. Asymmetric Schottky Contacts in Bilayer MoS$_2$ Field Effect Transistors. *Adv. Funct. Mater.* **2018**, *28*, 1800657.

(75) Di Bartolomeo, A.; Grillo, A.; Urban, F.; Iemmo, L.; Giubileo, F.; Luongo, G.; Amato, G.; Croin, L.; Sun, L.; Liang, S.-J.; Ang L. K. Asymmetric Schottky Contacts in Bilayer MoS$_2$ Field Effect Transistors. *Adv. Funct. Mater.* **2018**, *28*, 1800657.

(76) Grillo, A.; Di Bartolomeo, A. A Current–Voltage Model for Double Schottky Barrier Devices. *Adv. Electron. Mater.* **2020**, *7*, 2000979.

(77) Sze, S. M. *Physics of semiconductor devices*, 2nd ed., Wiley, New York, 1981.

(78) Di Bartolomeo, A.; Giubileo, F.; Luongo, G.; Iemmo, L.; Martucciello, N.; Niu, G.; Fraschke, M.; Skibitzki, O.; Schroeder, T.; Lupina, G. Tunable Schottky Barrier and High Responsivity in Graphene/Si-Nanotip Optoelectronic Device. *2D Mater.* **2017**, *4*, 015024.

(79) Cheiwchanchamnangij, T.; Lambrecht, W. R. L. Quasiparticle Band Structure Calculation of Monolayer, Bilayer, and Bulk MoS$_2$. *Phys. Rev. B* **2012**, *85*, 205302.

(80) Giannazzo, F.; Fisichella, G.; Piazza, A.; Agnello, S.; Roccaforte, F. Nanoscale Inhomogeneity of the Schottky Barrier and Resistivity in MoS$_2$ Multilayers. *Phys. Rev. B* **2015**, *92*, 081307(R).

(81) Son, Y.; Wang, Q. H.; Paulson, J. A.; Shih, C.-J.; Rajan, A. G.; Tvrdy, K.; Kim, S.; Alfeeli, B.; Braatz, R. D.; Strano, M. S. Layer Number Dependence of MoS$_2$ Photoconductivity Using Photocurrent Spectral Atomic Force Microscopic Imaging. *ACS Nano* **2015**, *9*, 2843-2855.

(82) Kim, C.; Moon, I.; Lee, D.; Choi, M. S.; Ahmed, F.; Nam, S.; Cho, Y.; Shin, H.-J.; Park, S.; Yoo, W. J. Fermi Level Pinning at Electrical Metal Contacts of Monolayer Molybdenum Dichalcogenides. *ACS Nano* **2017**, *11*, 1588-1596.

(83) Vaknin, Y.; Dagan, R.; Rosenwaks, Y. Schottky Barrier Height and Image Force Lowering in Monolayer MoS$_2$ Field Effect Transistors. *Nanomaterials* **2020**, *10*, 2346.

(84) Loth, S.; Wenderoth, M.; Ulbrich, R. G.; Malzer, S.; Döhler, G. H. Connection of Anisotropic Conductivity to Tip-Induced Space-Charge Layers in Scanning Tunneling Spectroscopy of p-doped GaAs. *Phys. Rev. B* **2007**, *76*, 235318.

(85) Melitz, W.; Shen, J.; Kummel, A. C.; Lee, S. Kelvin Probe Force Microscopy and its Application. *Surf. Sci. Rep.* **2011**, *66*, 1-27.

(86) Fernández Garrillo, P. A.; Grévin, B.; Chevalier, N.; Borowik, L. Calibrated Work Function Mapping by Kelvin Probe Force Microscopy. *Rev Sci Instrum.* **2018**, *89*, 043702.

(87) Gong, C.; Colombo, L.; Wallace, R. M.; Cho, K. The Unusual Mechanism of Partial Fermi Level Pinning at Metal-MoS$_2$ Interfaces. *Nano Lett.* **2014**, *14*, 1714-1720.

(88) Guo, Y.; Liu, D.; Robertson, J. Chalcogen Vacancies in Monolayer Transition Metal Dichalcogenides and Fermi Level Pinning at Contacts. *Appl. Phys. Lett.* **2015**, *106*, 173106.

(89) Gong, C.; Zhang, H.; Wang, W.; Colombo, L.; Wallace, R. M.; Cho, K. Band Alignment of Two-Dimensional Transition Metal Dichalcogenides: Application in Tunnel Field Effect Transistors. *Appl. Phys. Lett.* **2013**, *103*, 053513.

(90) Gong, C.; Zhang, H.; Wang, W.; Colombo, L.; Wallace, R. M.; Cho, K. Band Alignment of Two-Dimensional Transition Metal Dichalcogenides: Application in Tunnel Field Effect Transistors. *Appl. Phys. Lett.* **2013**, *103*, 053513.

(91) Park, J. H.; Sanne, A.; Guo, Y.; Amani, M.; Zhang, K.; Movva, H. C. P.; Robinson, J. A.; Javey, A.; Robertson, J.; Banerjee, S. K.; Kummel, A. C. Defect Passivation of Transition Metal Dichalcogenides via a Charge Transfer van der Waals Interface. *Sci. Adv.* **2017**, *3*, e1701661.

(92) Lee, H.; Deshmukh, S.; Wen, J.; Costa, V. Z.; Schuder, J. S.; Sanchez, M.; Ichimura, A. S.; Pop, E.; Wang, B.; Newaz, A. K. M. Layer-Dependent Interfacial Transport and Optoelectrical Properties of MoS$_2$ on Ultraflat Metals. *ACS Appl. Mater. Interfaces* **2019**, *11*, 31543-31550.

(93) Park, S.; Schultz, T.; Shin, D.; Mutz, N.; Aljarb, A.; Kang, H. S.; Lee, C.-H.; Li, L.-J.; Xu, X.; Tung, V. et al. The Schottky–Mott Rule Expanded for Two-Dimensional Semiconductors: Influence of Substrate Dielectric Screening. *ACS Nano* **2021,** *15*, 14794-14803.

(94) Zhao, Y.; Tripathi, M.; Čerņevičs, K.; Avsar, A.; Ji, H. G.; Gonzalez Marin, J. F.; Cheon, C.-Y.; Wang, Z.; Yazyev, O. V.; Kis, A. Electrical Spectroscopy of Defect States and Their Hybridization in Monolayer MoS$_2$. *Nat. Commun.* **2023**, *14*, 44.

(95) Nijhuis, C. A.; Reus, W. F.; Whitesides, G. M. Molecular Rectification in Metal−SAM−Metal Oxide−Metal Junctions. *J. Am. Chem. Soc.* **2009**, *131*, 17814-17827.

(96) Trasobares, J.; Rech, J.; Jonckheere, T.; Martin, T.; Aleveque, O.; Levillain, E.; Diez-Cabanes, V.; Olivier, Y.; Cornil, J.; Nys, J. P. et al. Estimation of π-π Electronic Couplings from Current Measurements. *Nano Lett.* **2017**, *17*, 3215–3224.

(97) QuantumATK T-2022.03, https://www.synopsys.com/manufacturing/quantumatk/atomistic-simulation-products.html.

(98) Smidstrup, S.; Markussen, T.; Vancraeyveld, P.; Wellendorff, J.; Schneider, J.; Gunst, T.; Verstichel, B.; Stradi, D.; Khomyakov, P. A.; Vej-Hansen, U. G.; et al. QuantumATK: an Integrated Platform of Electronic and Atomic-Scale Modelling Tools. *J. Phys.: Condens. Matter* **2020**, *32*, 015901.

(99) Perdew, J. P.; Burke, K.; Ernzerhof, M. Generalized Gradient Approximation Made Simple. *Phys. Rev. Lett.* **1996**, *77*, 3865-3868.

(100) Becke, A. D. A New Mixing of Hartree–Fock and Local Density-Functional Theories. *J. Chem. Phys.* **1993**, *98*, 1372-1377.

(101) Heyd, J.; Scuseria, G. E.; Ernzerhof, M. Hybrid Functionals Based on a Screened Coulomb Potential. *J. Chem. Phys.* **2003**, *118*, 8207-8215.

(102) Krukau, A. V.; Vydrov, O. A.; Izmaylov, A. F.; Scuseria, G. E. Influence of the Exchange Screening Parameter on the Performance of Screened Hybrid Functionals. *J. Chem. Phys.* **2006**, *125*, 224106.

(103) Troullier, N.; Martins, J. L. Efficient Pseudopotentials for Plane-Wave Calculations. *Phys. Rev. B* **1991**, *43*, 1993-2006.

(104) Premasiri, K.; Gao, X. P. A. Tuning Spin-Orbit Coupling in 2D Materials for Spintronics: a Topical Review. *J. Phys. Condens. Matter.* **2019**, *31*, 193001.

(105) Yang, B.; Tu, M.-F.; Kim, J.; Wu, Y.; Wang, H.; Alicea, J.; Wu, R.; Bockrath, M.; Shi, J. Tunable Spin–Orbit Coupling and Symmetry-Protected Edge States in Graphene/$WS_2$. *2D Mater.* **2016**, *3*, 031012.

(106) Slassi, A.; Gali, A.; Cornil, J.; Pershin, A. Non-covalent Functionalization of Pristine and Defective $WSe_2$ by Electron Donor and Acceptor Molecules. *ACS Appl. Electron. Mater.* **2023**, *5*, 1660-1669.

(107) Sohn, A.; Moon, H.; Kim, J.; Seo, M.; Min, K.-A.; Lee, S. W.; Yoon, S.; Hong, S.; Kim, D.-W. Band Alignment at Au/$MoS_2$ Contacts: Thickness Dependence of Exfoliated Flakes. *J. Phys. Chem. C* **2017**, *121*, 22517-22522.

(108) Lattyak, C.; Gehrke, K.; Vehse, M. Layer-Thickness-Dependent Work Function of $MoS_2$ on Metal and Metal Oxide Substrates. *J. Phys. Chem. C* **2022**, *126*, 13929-13935.

(109) Markeev, P. A.; Najafidehaghani, E.; Gan, Z.; Sotthewes, K.; George, A.; Turchanin, A.; de Jong, M. P. Energy-Level Alignment at Interfaces Between Transition-Metal Dichalcogenide

Monolayers and Metal Electrodes Studied with Kelvin Probe Force Microscopy. *J. Phys. Chem. C* **2021**, *125*, 13551-13559.

(110) Neaton, J. B.; Hybertsen, M. S.; Louie, S. G. Renormalization of Molecular Electronic Levels at Metal-Molecule Interfaces. *Phys. Rev. Lett.* **2006**, *97*, 216405.

*Supporting Information*

___________________________________________________________________________

**Electronic Properties of Electroactive Ferrocenyl-Functionalized MoS$_2$**

Trung Nghia Nguyên Lê,[a] Kirill Kondratenko,[b] Imane Arbouch,[c] Alain Moréac,[d] Jean-Christophe Le Breton,[d] Colin van Dyck,[e] Jérôme Cornil,[c] Dominique Vuillaume[b] and Bruno Fabre[a],*

[a] *CNRS, ISCR (Institut des Sciences Chimiques de Rennes)-UMR6226, Univ Rennes, Rennes F-35000, France.*

[b] *Institute for Electronics Microelectronics and Nanotechnology (IEMN), CNRS, University of Lille, Av. Poincaré, Villeneuve d'Ascq, France.*

[c] *Laboratory for Chemistry of Novel Materials, University of Mons, B-7000 Mons, Belgium*

[d] *CNRS, IPR (Institut de Physique de Rennes)-UMR 6251, Univ Rennes, Rennes F-35000, France.*

[e] *Theoretical Chemical Physics group, University of Mons, B-7000 Mons, Belgium.*

*\*E-mail: bruno.fabre@univ-rennes1.fr*

*Supporting Information*
______________________________________________________________________

## 1. Preparation and characterization of the MoS$_2$ samples

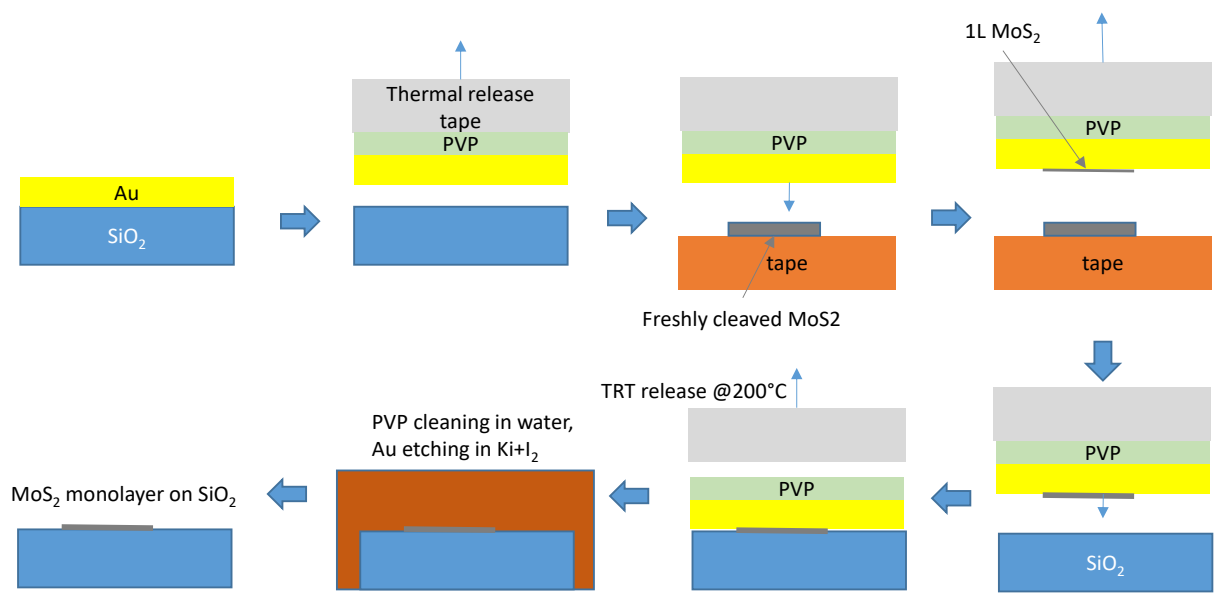

**Figure S1.** Preparation of MoS$_2$ monolayer-coated SiO$_2$/Si surfaces.

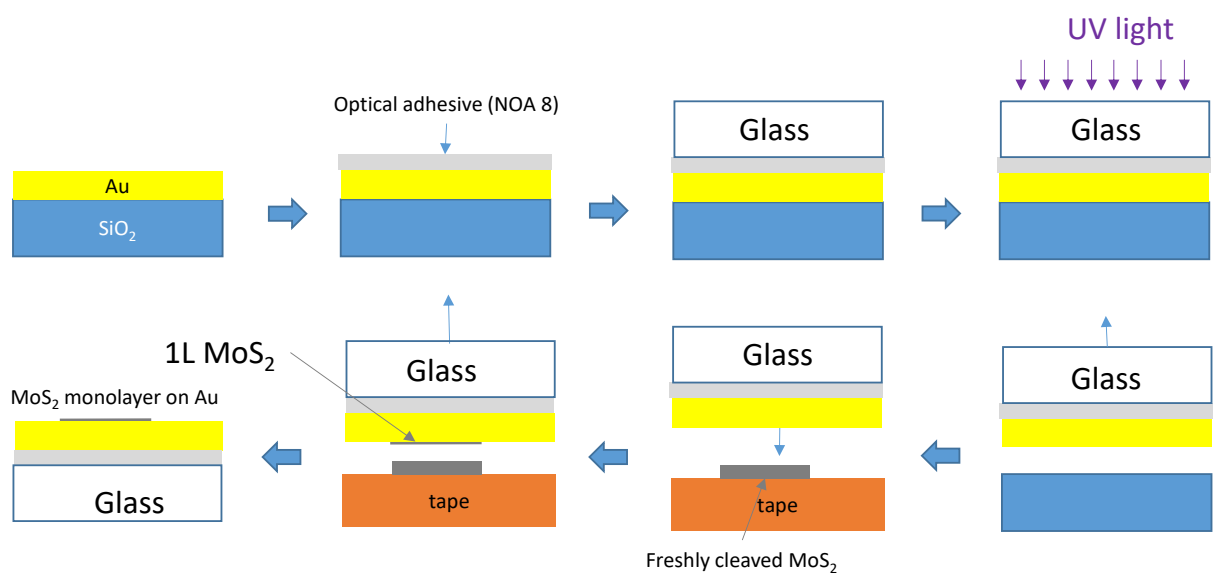

**Figure S2.** Preparation of MoS$_2$ monolayer-coated Au surfaces.

S3

## 2. Monitoring of each washing cycle by ¹H-NMR spectroscopy

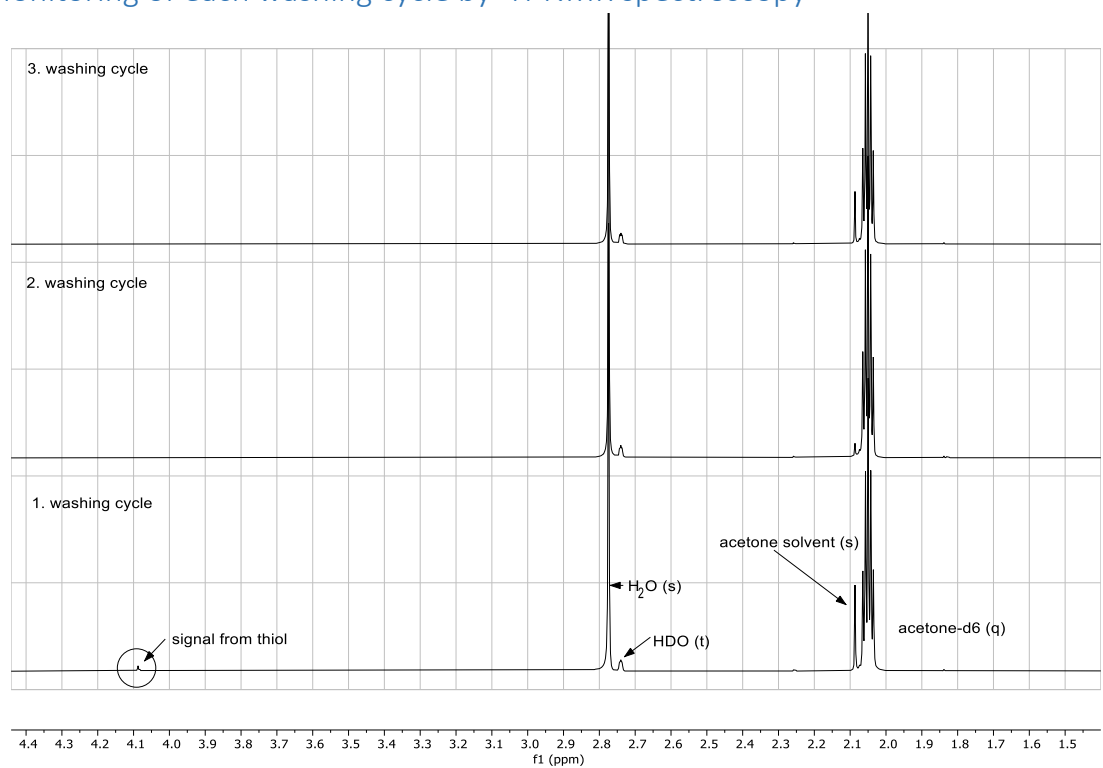

**Figure S3.** $^1$H-NMR spectra of the residues after each washing cycle of the 6-(ferrocenyl)hexanethiol-grafted defective MoS$_2$ surface with acetone-d$_6$. Traces of thiol could only be observed after the first cycle. The following washing cycles showed that the majority of the physisorbed species has been removed.

*In order to remove physisorbed species after ferrocene grafting, the sample was washed by immersing it into multiple acetone baths. For each washing cycle, the solvent was collected and subsequently removed under high vacuum. The residues were dissolved again in acetone-d$_6$ and analyzed by $^1$H-NMR spectroscopy. The 6-(ferrocenyl)hexanethiol molecule could be observed after the first washing cycle, specifically around 4.1 ppm, as depicted in **Erreur ! Source du renvoi introuvable.**. After two additional washing cycles, no traces of the thiol could be detected indicating that the majority of the physisorbed species was removed. Therefore, we can conclude that the electrochemical response observed by cyclic voltammetry was attributed to mainly covalently bonded molecules to the surface.*

## 3. Influence of the absence of plasma treatment

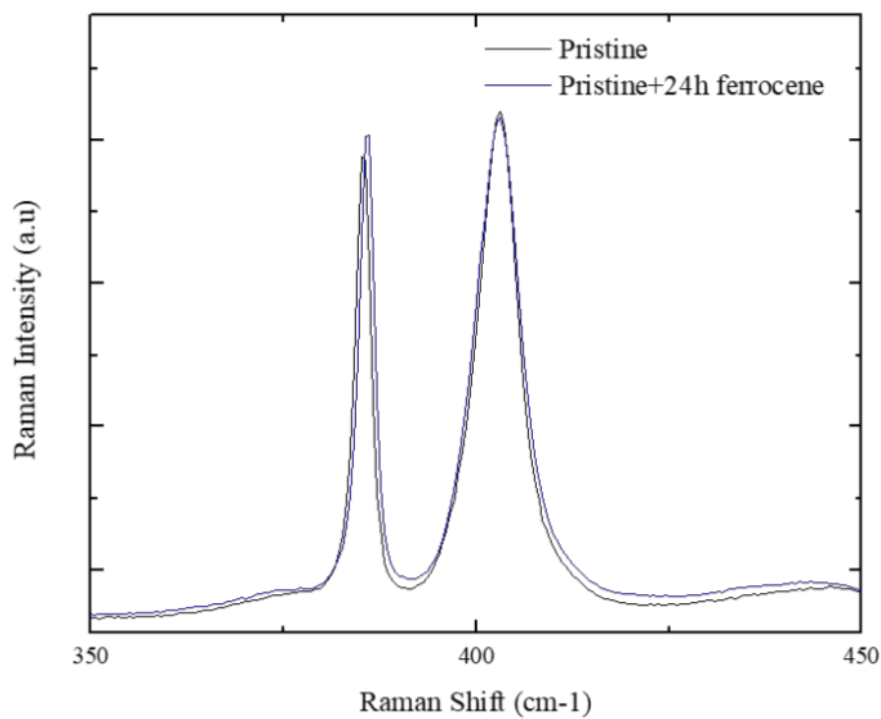

**Figure S4.** Raman spectra of pristine MoS$_2$ (without plasma treatment) before and after 24 h in the 6-(ferrocenyl)hexanethiol grafting solution.

*Figure S4 shows the Raman spectra of a pristine MoS$_2$ sample before and after immersion for 24 h in the grafting solution. For both samples, we observe the E' and A'$_1$ peaks with a similar lineshape. A small blueshift of 0.43 cm$^{-1}$ of the E' peak is observed which is indicative of a slight strain probably caused by the presence of adsorbed molecules and solvent residues. The A' peak on the other side remains at the same position.*

## 4. Fitting of Raman spectra

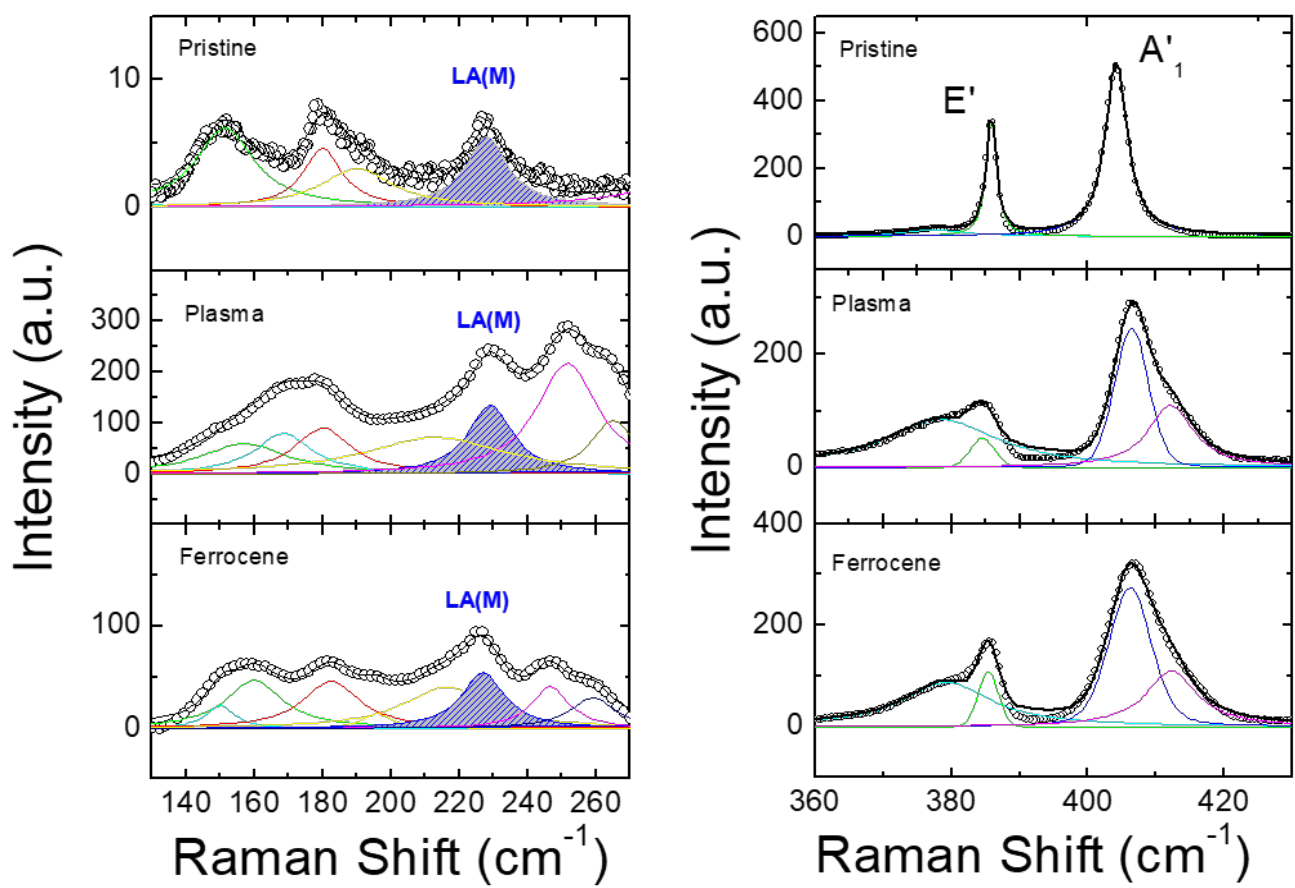

**Figure S5.** Fitting results of Raman spectra after background subtraction for pristine, plasma-treated and ferrocene-grafted MoS$_2$ monolayer in the LA(M) (left) and the E' and A'$_1$ (right) regions.

## 5. AFM measurements

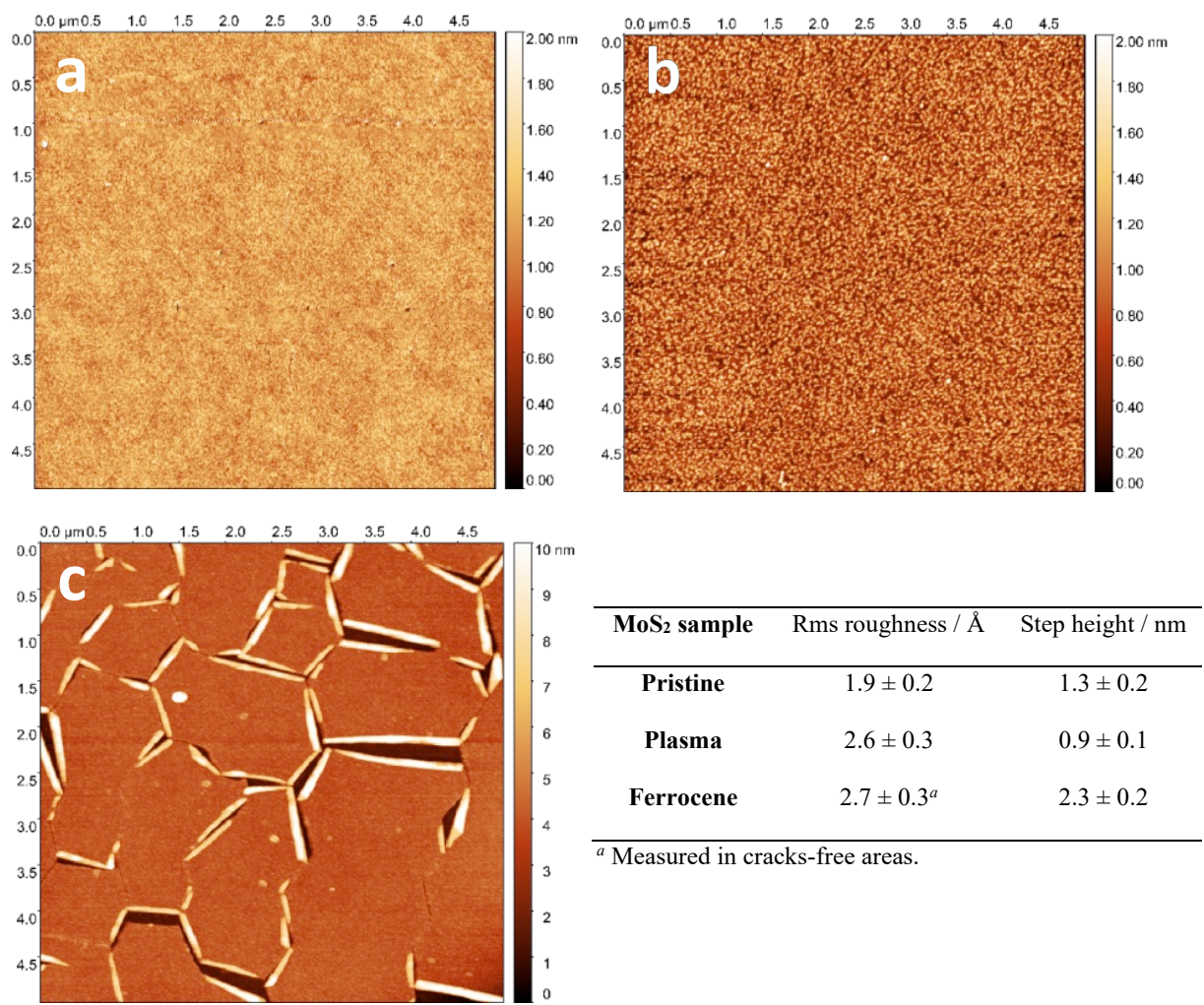

**Figure S6.** AFM images (5 x 5 $\mu m^2$) of pristine (a), plasma-treated (b) and ferrocene-grafted (c) $MoS_2$ deposited on $SiO_2/Si$. (Bottom right) Root-mean-square roughness and step height determined by AFM (see Fig. 4 in the main text).

| $MoS_2$ sample | Rms roughness / Å | Step height / nm |
|---|---|---|
| **Pristine** | 1.9 ± 0.2 | 1.3 ± 0.2 |
| **Plasma** | 2.6 ± 0.3 | 0.9 ± 0.1 |
| **Ferrocene** | 2.7 ± 0.3[a] | 2.3 ± 0.2 |

[a] Measured in cracks-free areas.

6. Cyclic voltammetry measurements

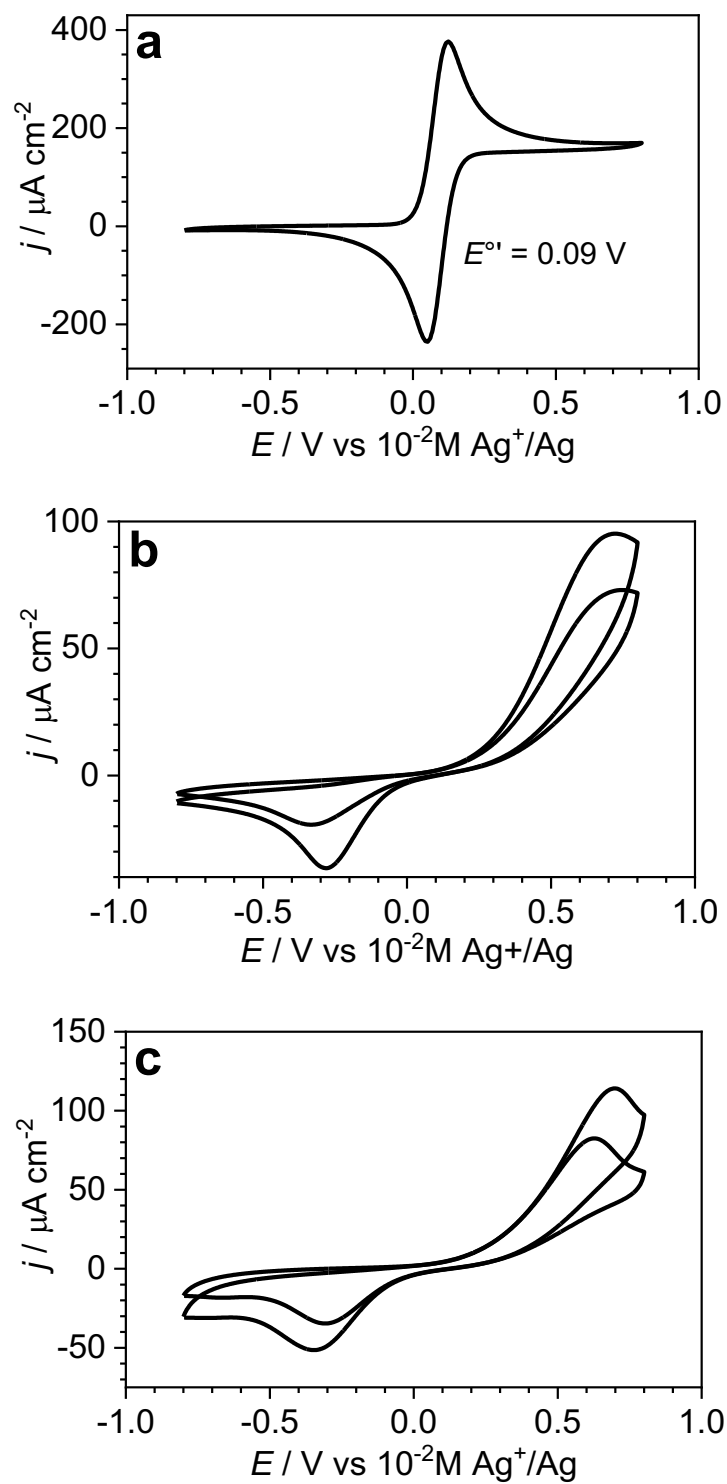

**Figure S7.** Cyclic voltammograms of ferrocene (1 mM) at a 1 mm-diameter platinum electrode at 80 mV s$^{-1}$ (a), SiO$_x$/ $p^{++}$Si(100) (b) and MoS$_2$-coated SiO$_x$/ $p^{++}$Si(100) (c) at 20 and 40 mV

s⁻¹, respectively. Electrolyte: $CH_3CN$ + 0.1M $Bu_4NClO_4$. The peak-to-peak separation $\Delta E_p$ is 70 mV (a), increasing to 980 mV (b) and 930 mV (c) which reflects the decrease in the electron transfer kinetics caused by the insulating $SiO_x$ layer.

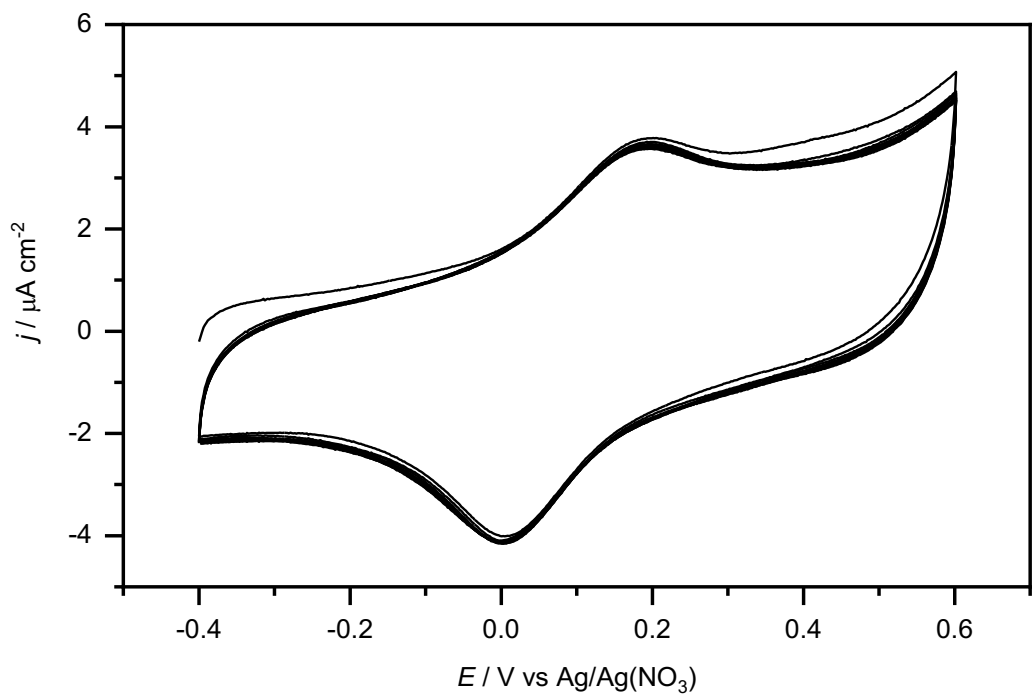

**Figure S8.** Cyclic voltammograms at 100 mV s⁻¹ of defective $MoS_2$-coated $SiO_x$ /p⁺⁺Si(100) surface after immersion for 24 h into an acetone solution containing (ferrocenyl)hexanethiol at 1 mM over the course of 10 successive scans. Electrolyte: $CH_3CN$ + 0.1M $Bu_4NClO_4$.

*As depicted in Figure S8, cyclic voltammograms were recorded at 100 mV s⁻¹ in CH₃CN medium over the course of 10 successive scans. Since no significant changes, e.g. loss of electrochemical signal, were observed, a strong interaction between the grafted electroactive molecules and surface was concluded.*

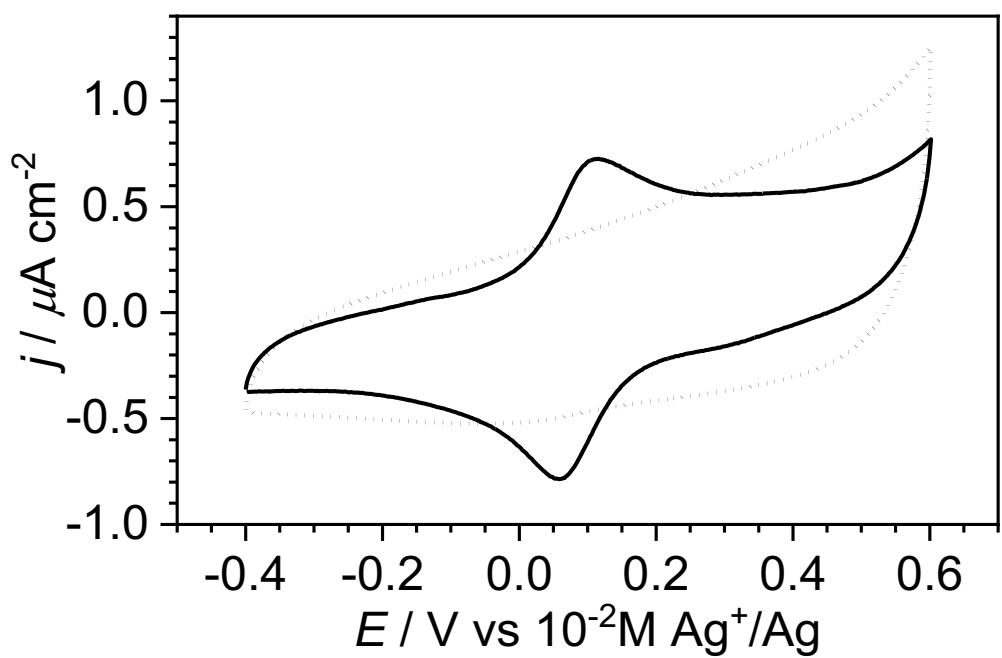

**Figure S9.** Cyclic voltammograms at 0.02 V s$^{-1}$ in CH$_3$CN + 0.1 M Bu$_4$NClO$_4$ of bare SiO$_x$/$p^{++}$Si(100) (dotted line) and defective MoS$_2$-coated SiO$_x$/$p^{++}$Si(100) (solid line) surfaces after immersion for 24 h into an acetone solution containing 6-(ferrocenyl)hexanethiol at 1 mM.

## 7. Effective electron mobility measurements

The mobility in linear regime ($V_D < V_G$) is calculated as usual by:

$$\mu_{eff} = \frac{L}{W} \frac{\partial I_D/\partial V_G}{C_{ox} |V_D|}$$

(S1)

with $\partial I_D/\partial V_G$ the slope of the linear part of the curve, $C_{ox}$ the SiO$_2$ capacitance (11.5 nF/cm$^2$ for a 300 nm-thick thermal SiO$_2$), $V_D$ the drain voltage (in absolute value), $L$ and $W$ the channel length and width, respectively, for a rectangular channel between the two source and drain electrodes. The length $L$ is given by the position of the C-AFM with respect to the edge of the lithographed electrode. The width $W$ is more difficult to estimate since the channel has a trapezoidal shape defined, at one end, by the size of the MoS$_2$ flake at the contact with the lithographed electrode and, at the other end, the size of the C-AFM contact. Typical optical images are shown in Figure S10. For simplicity, we consider that $W$ is restricted to the more direct electronic pathway between the C-AFM tip and the electrode (i.e. marked by the rectangular red dot line in Figure S10d). We assume that $W$ is of the same order of magnitude as the nominal radius of the tip, $W \approx 20$ nm. This hypothesis clearly overestimates the calculated mobility, which is used here only for the purpose of comparison between the samples measured under the same conditions and must be used with caution for comparison with data reported in the literature. A more precise estimation of the mobility for this strongly asymmetric electrode configuration would require 2D device simulations (which is out of the scope of this work). Table S1 summarizes the data for all the measured samples.

*Supporting Information*

___

**Table S1**. Calculated effective mobility ($\mu_{\text{eff}}$), slope of the transfer characteristic ($\partial I_D/\partial V_G$), applied drain voltage $V_D$ and estimated channel length ($L$) for the different MoS$_2$ samples.

| Sample | $\partial I_D/\partial V_G$ (A/V) | $V_D$ (V) | $L$ (μm) | $\mu_{\text{eff}}$ (cm$^2$ V$^{-1}$ s$^{-1}$) |
|---|---|---|---|---|
| **Pristine MoS$_2$** | | | | |
| #1 | 9.6 x 10$^{-11}$ | -3 | 20 | 2.8 |
| #2 | 1.7 x 10$^{-9}$ | -3 | 8 | 22.8 |
| #3 | 2.8 x 10$^{-10}$ | -5 | 20 | 4.5 |
| #4 | 2 x 10$^{-9}$ | -5 | 12 | 20.9 |
| #5 | 6.4 x 10$^{-10}$ | -5 | 16 | 8.9 |
| **Ferrocene MoS$_2$** | | | | |
| #1 | 8 x 10$^{-11}$ | -4 | 4 | 0.35 |
| #2 | 2.2 x 10$^{-10}$ | -5 | 6 | 1.1 |
| #3 | 4.4 x 10$^{-10}$ | -3 | 4 | 2.6 |

## 8. Optical images of the samples for the in-plane electron transfer (ET) measurements

We present several optical images (taken from the camera of the C-AFM equipment) used to estimate the channel length ($L$), the width ($W$) -*vide supra*- and the contact area between the MoS$_2$ flake and the lithographed electrode. $L$ was in the range 4 to 20 $\mu$m. The contact area between the lithographed Au electrode and the MoS$_2$ flake ($S_1$ in eq. 1, main text) was typically $3 \times 10^3 – 2 \times 10^4$ $\mu m^2$.

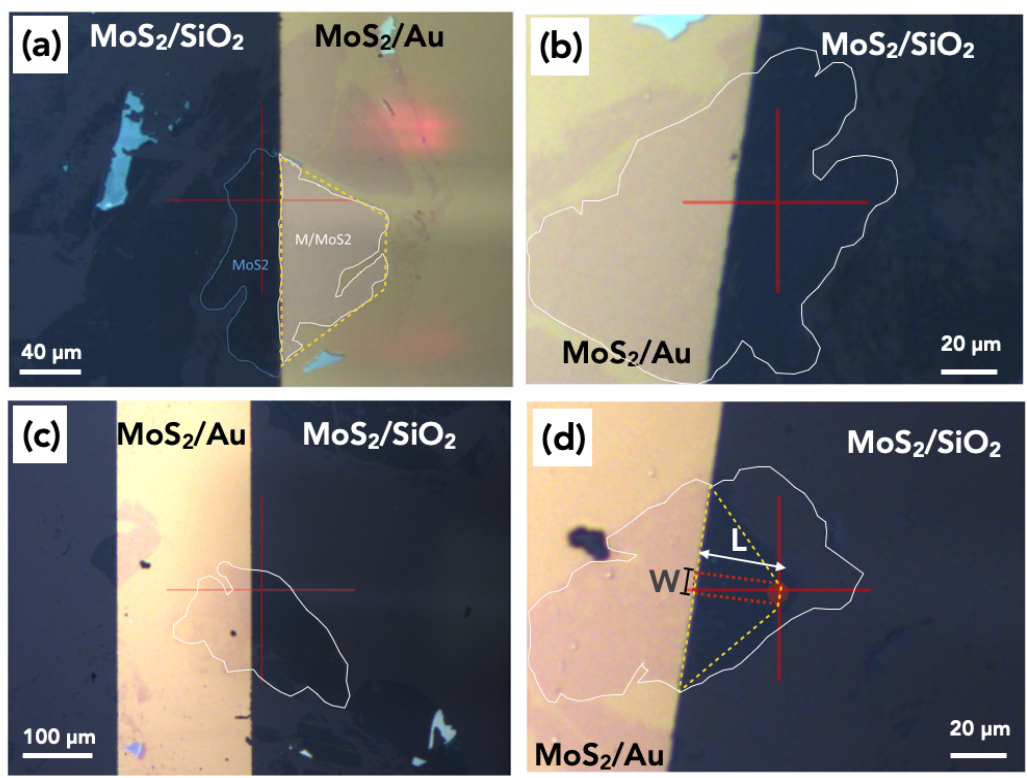

**Figure S10**. Optical images of several MoS$_2$ flakes connected to lithographed electrodes. The red crosses indicate the position of the C-AFM tip. The measured flakes are delineated by the full thin white lines. The contact area between the lithographed Au electrode and the MoS$_2$ flake ($S_1$ in eq. 1, main text) is estimated from the area of a "close" simple geometrical shape (e.g. a trapeze, yellow dashed line in the case shown in the panel (a)). The transistor channel length $L$ and width $W$ used to calculate the effective mobility are estimated as indicated in the panel (d) by the red dashed line. $L$ is the distance between the C-AFM tip and the edge of the Au electrode, the effective width *W* is given by the diameter of the tip (red spot, not on scale), while the real shape of the transistor channel is indicated by the yellow dashed line in the panel (d).

## 9. Estimation of the contact area for the electronic transport measurements

As usually reported in literature[1,2,3,4] for C-AFM measurements, the contact radius, $r_c$, between the C-AFM tip and the surface, and the film elastic deformation, $\delta$, are estimated from a Hertzian model:[5]

$$r_c^2 = \left(\frac{3RF}{4E^*}\right)^{2/3} \tag{S1}$$

$$\delta = \left(\frac{9}{16R}\right)^{1/3}\left(\frac{F}{E^*}\right)^{2/3} \tag{S2}$$

with $F$ the tip loading force (15 nN), $R$ the tip radius (20 nm) and $E^*$ the reduced effective Young modulus defined as:

$$E^* = \left(\frac{1}{E_s^*} + \frac{1}{E_{tip}^*}\right)^{-1} = \left(\frac{1-v_s^2}{E_s} + \frac{1-v_{tip}^2}{E_{tip}}\right)^{-1} \tag{S3}$$

In this equation, $E_{s/tip}$ and $v_{s/tip}$ are the Young modulus and the Poisson ratio of the sample (MoS$_2$ monolayer) and C-AFM tip, respectively. For the Au tip, we have $E_{tip}$ = 78 GPa. For a MoS$_2$ monolayer, we used $E_s$=270 GPa.[6,7] The Poisson ratio is $v_{tip}$=0.44, $v_s$=0.125.[8] Using these parameters, we estimate a contact area of ≈ 7 nm² ($r_c$ ≈ 1.8 nm) and $\delta$ ≈ 0.1 nm. For a ferrocenyl monolayer, which is softer than MoS$_2$, the parameters are not known and, in general, they are not easily determined in such a monolayer material. Thus, we consider the value of an effective Young modulus of the monolayer $E^*_{SAM}$ = 38 Gpa, as determined for the "model system" alkylthiol self-assembled monolayers (SAMs) from a combined mechanic and electron transport study.[3] We get a contact area of ≈ 13 nm² ($r_c$ ≈ 2 nm and $\delta$ ≈ 0.2 nm). For simplicity, we analyzed all the data with a mean contact area of ≈ 10 nm².

## 10. Complementary electronic transport data

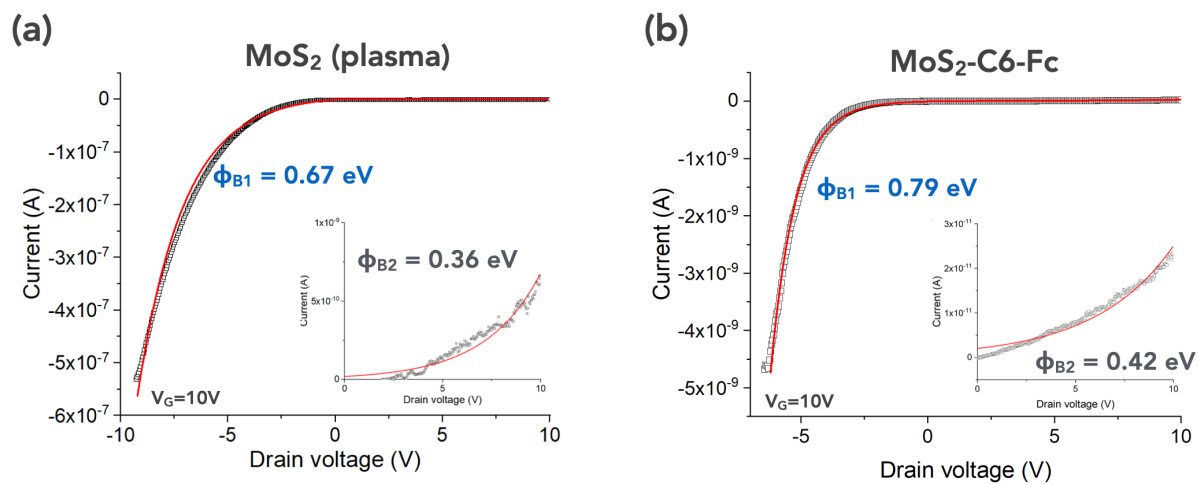

**Figure S11**. In-plane current vs. the drain voltage curves (black dots) and fits with eq. 1 (red lines) for (a) the plasma-treated MoS$_2$ and (b) the ferrocene-grafted MoS$_2$ samples. The insets show the $V > 0$ region. The extracted SBHs $\phi_{B1}$ and $\phi_{B2}$ are marked in the panels.

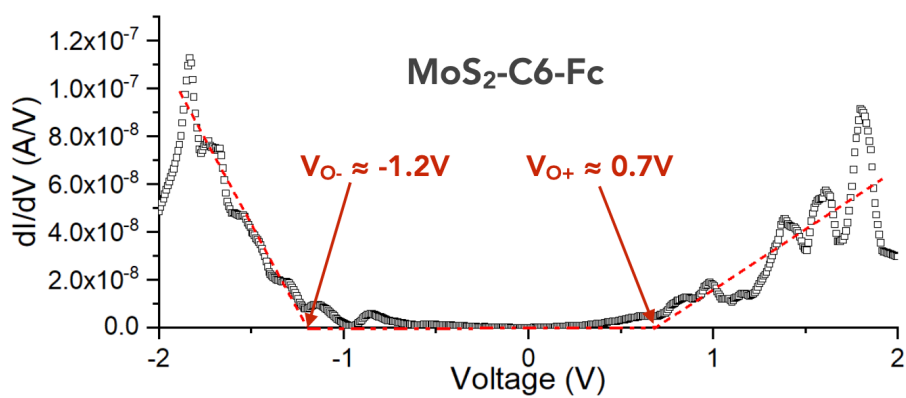

**Figure S12**. First derivative $\partial I/\partial V$ - $V$ (numerical derivation from the mean $\bar{I} - V$ curve) of the ferrocene-grafted MoS$_2$ sample.

## 11. UPS

Figure S13 shows the valence band edge measured by UPS on the pristine and plasma-treated MoS$_2$. The valence band edges are determined by the intercept of the background and the tail of DoS (dotted lines) at 1.34 and 1.13 eV, respectively. For the ferrocene-grafted MoS$_2$ (Fig. S13c), we do not detect the HOMO peak of neutral Fc in the MoS$_2$ band gap (as observed for SAM of alkylthiol-Fc directly grafted on Au),[9] which is consistent with a majority of the Fc moieties in the oxidized Fc$^+$ states, the HOMO of Fc$^+$, at larger binding energy, being mixed with the MoS$_2$ VB signal.

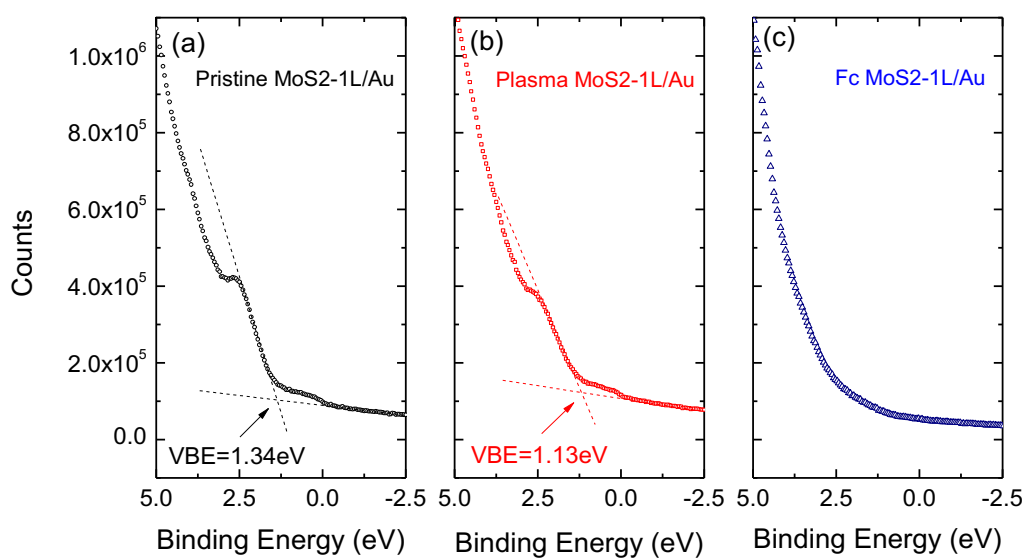

**Figure S13**. UPS of the (a) pristine, (b) plasma-treated and (c) ferrocene-grafted MoS$_2$ samples near the Fermi level. The intercepts of the dashed lines are used to define the valence band edge of MoS$_2$.

*Supporting Information*

________________________________________________________________

12. MoS$_2$/Au structures with ghost atoms (DFT calculations)

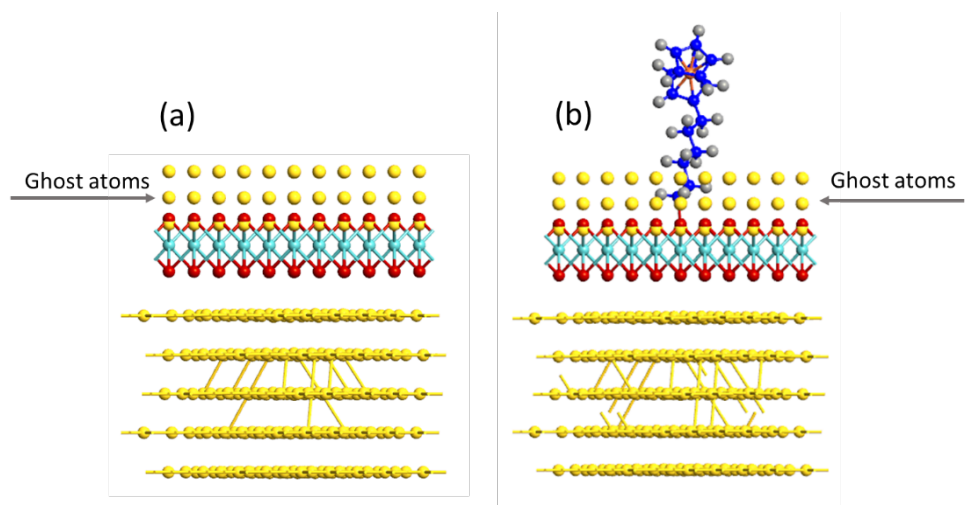

**Figure S14**. A side view of the MoS$_2$/Au structures when adding three layers of gold ghost atoms for (a) pristine MoS$_2$/Au and (b) ferrocene-grafted MoS$_2$/Au. The ghost atoms were created by copying the three layers of MoS$_2$ at a distance of 3 Å from the bottom sulfur layer of MoS$_2$ and then converting the sulfur and molybdenum atoms to gold ghost atoms.

S18

## 13. References

(1) Cui, X. D.; Primak, A.; Zarate, X.; Tomfohr, J.; Sankey, O. F.; Moore, A. L.; Moore, T. A.; Gust, D.; Harris, G.; Lindsay, S. M. Reproducible Measurement of Single-Molecule Conductivity. *Science* **2001**, *294*, 571-574.

(2) Cui, X. D.; Zarate, X.; Tomfohr, J.; Sankey, O. F.; Primak, A.; Moore, A. L.; Moore, T. A.; Gust, D.; Harris, G.; Lindsay, S. M. Making Electrical Contacts to Molecular Monolayers. *Nanotechnology* **2002**, *13*, 5-14.

(3) Engelkes, V. B.; Frisbie, C. D. Simultaneous Nanoindentation and Electron Tunneling through Alkanethiol Self-Assembled Monolayers. *J. Phys. Chem. B* **2006**, *110*, 10011-10020.

(4) Morita, T.; Lindsay, S. Determination of Single Molecule Conductances of Alkanedithiols by Conducting-Atomic Force Microscopy with Large Gold Nanoparticles. *J Am. Chem. Soc.* **2007**, *129*, 7262-7263.

(5) Johnson, K. L. *Contact Mechanics*, Cambridge University Press, New York, 1987.

(6) Bertolazzi, S.; Brivio, J.; Kis, A. Stretching and Breaking of Ultrathin $MoS_2$. *ACS Nano* **2011**, *5*, 9703-9709.

(7) Li, Y.; Yu, C.; Gan, Y.; Jiang, P.; Yu, J.; Ou, Y.; Zou, D.-F.; Huang, C.; Wang, J.; Jia, T.; Luo, Q.; Yu, X.-F.; Zhao, H.; Gao, C.-F.; Li, J. Mapping the Elastic Properties of Two-Dimensional $MoS_2$ via Bimodal Atomic Force Microscopy and Finite Element Simulation. *npj Computational Materials* **2018**, *4*, 49.

(8) Son, Y.; Wang, Q. H.; Paulson, J. A.; Shih, C.-J.; Rajan, A. G.; Tvrdy, K.; Kim, S.; Alfeeli, B.; Braatz, R. D.; Strano, M. S. Layer Number Dependence of $MoS_2$ Photoconductivity Using Photocurrent Spectral Atomic Force Microscopic Imaging. *ACS Nano* **2015**, *9*, 2843-2855.

(9) Wong, R. A.; Yokota, Y.; Wakisaka, M.; Inukai, J.; Kim, Y. Discerning the Redox-Dependent Electronic and Interfacial Structures in Electroactive Self-Assembled Monolayers. *J. Am. Chem. Soc.* **2018**, *140*, 13672-13679.